\documentclass[12pt]{article}
\usepackage[T1]{fontenc}
\usepackage{fancyhdr}
\usepackage{amsmath,amsthm,amsfonts,amssymb}
\usepackage{epic,eepic}
\usepackage{graphicx}
\usepackage{authblk}
\usepackage[active]{srcltx}

\graphicspath{{Figures/}}

\usepackage{color}

\usepackage{float}%
\floatstyle{plaintop}%
\restylefloat{table}
\usepackage[font={small,it}]{caption} 
\usepackage{fullpage}
\usepackage{algpseudocode}
\usepackage{algorithm}
\usepackage{setspace}

\usepackage[style=authoryear,hyperref,backend=bibtex,maxcitenames=2,maxbibnames=99,dashed=false]{biblatex}
\bibliography{biblio}

\theoremstyle{plain}

\theoremstyle{definition}

\theoremstyle{remark}

\numberwithin{equation}{section} 
\numberwithin{figure}{section} 
\numberwithin{theorem}{section}

\DeclareMathOperator*{\argmax}{arg\,max}

\usepackage{hyperref}

\begin{document}
\begin{center}

\Large{\sf\bf
	Choosing the number of clusters in a finite mixture model using an exact Integrated Completed Likelihood criterion\\
}
{
	\vspace{0.05cm} Marco Bertoletti$^{\star}$, Nial Friel$^{\dagger}$ and Riccardo Rastelli$^{\dagger}$\footnote{Corresponding author: riccardo.rastelli@ucdconnect.ie}\\
 	{\footnotesize 
	\vspace{0.3cm}
	
	$^{\star}$Dept. of Statistical Sciences, University of Bologna, Italy\\
	
	$^{\dagger}$Insight: Centre for Data Analytics and School of Mathematical Sciences, University College Dublin, Ireland}
}

\end{center}

\section*{Abstract}
The ICL criterion has proven to be a very popular approach in Model-Based Clustering through automatically choosing the number of clusters in a mixture model.
This approach effectively maximises the complete data likelihood, thereby including the allocation of observations to clusters in the model selection
criterion. However for practical implementation one needs to introduce an approximation in order to estimate the ICL. Our contribution here is to illustrate 
that through the use of conjugate priors one can derive an exact expression for ICL and so avoiding any approximation. Moreover, we illustrate how one 
can find both the number of clusters and the best allocation of observations in one algorithmic framework. The performance of our algorithm
is presented on several simulated and real examples.\\

\noindent
{\bf Keywords:} Integrated Completed Likelihood, Finite mixture models, Model-based clustering, Greedy search.

\section{Introduction}\label{sec:Introduction}

Finite mixture models are a widely used approach for parametric cluster analysis. Choosing the number of components in the mixture model, usually
viewed as a model choice problem, is a crucial issue.
In a Bayesian framework, model choice can be dealt with in a Markov Chain Monte Carlo framework where the number of components is estimated simultaneously with 
model parameters using the Reversible Jump algorithm of \textcite{green1995reversible}, extended to the context of finite mixtures by \textcite{richardson:green97}. 
An alternative approach has been introduced in \textcite{nobile2007bayesian}, where authors propose to integrate out model parameters, in order to achieve better 
estimation and more efficient sampling. The resulting algorithm, called the Allocation Sampler, carries out inference on the allocations (cluster labels 
of the observations) and the number of components $K$ in one framework. Similar approaches based on collapsing model parameters have been applied to different contexts, 
such as network models, as shown in \textcite{wyse2012block,wyse2014inferring,come2013model,mcdaid2013improved}.

A more standard approach to model choice relies instead on the maximization of the Bayesian Information Criterion (BIC), which has again a Bayesian derivation but can be 
used in a more objective and frequentist fashion. 
Throughout the paper we will denote the observed data with $\textbf{x}$. Each observation is allocated in one group and one only,
and the so-called allocation variable $\textbf{z}$ is categorical and takes values in $\left\{1,\dots, K\right\}$. Now, let $\hat{\theta}_K$ be the estimated model parameters under 
the assumption of a model with $K$ components, then the log model evidence 
$\log f\left( \textbf{x} \middle\vert K\right)$ is approximated by the BIC, which is defined as:
\begin{equation}\label{BIC1}
 BIC(K) = \log f\left( \textbf{x}\middle\vert K,\hat{\textbf{z}},\hat{\theta}_K \right)- \frac{1}{2}\nu_{K}\log(n),
\end{equation} 
where $\hat{\textbf{z}}$ are the MLE allocations corresponding to the estimated parameters, while $n$ is the number 
of observations and $\nu_M$ the number of parameters under the assumption of $K$ groups. 
The BIC index approximates the $\log$ model evidence in that both the parameters and the allocations 
are replaced with maximum likelihood estimates. Hence, among all the possible choices, the model maximizing BIC is chosen.

However, an interesting variant of the BIC method has been introduced in \textcite{biernacki2000assessing}. 
In this work, the authors propose 
to base model choice on the Integrated Completed Likelihood (ICL) criterion, which differs from BIC in that the complete data (observations and allocations)
are used to make the decision concerning the model to be used.
More specifically, the ICL index turns out to be equal to the BIC penalised by the estimated mean entropy:
\begin{equation}\label{ICLdef1}
\begin{split}
 ICL(K) := BIC(K) - \sum_{g=1}^{K}\sum_{i=1}^{n} p\left( z_i\middle\vert \textbf{x},\hat{\theta}_K, K \right)\log p\left( z_i\middle\vert \textbf{x},\hat{\theta}_K, K \right).
\end{split}
\end{equation}
Using a similar reasoning to that of BIC, the ICL index approximates the complete $log$ model evidence, given by:
\begin{equation}\label{exactICL1}
\begin{split}
 \log f\left( \textbf{x},\textbf{z} \middle\vert K\right)
 =\log\int_{\boldsymbol{\Theta}_K}f\left( \textbf{x},\textbf{z} \middle\vert \boldsymbol{\theta}_K,K\right)\pi\left( \boldsymbol{\theta}_K \right)d\boldsymbol{\theta}_K.
\end{split}
\end{equation}
The performances of ICL compared to those of BIC in choosing the number of components in a model-based clustering framework have been assessed in several works
\parencite{steele2010performance,biernacki2000assessing,baudry2010combining,maitra2010assessing}.
The key difference between BIC and ICL is that the latter includes an additional term (the estimated mean entropy) that penalises clustering configurations exhibiting
overlapping groups: low-entropy solutions with well-separated groups are preferred to configurations that give the best match with regard to the distributional 
assumptions.
An interpretation of the two (possibly different) optimal solutions is described in \cite{baudry2010combining}: 
while BIC allows a very efficient estimation of the number of components for the mixture model, ICL estimates instead the number of clusters
appearing in the data. Indeed, the term cluster usually refers to a homogeneous cloud of points, making it feasible for a cluster to be 
made of several mixture components. Due to the extra penalization term, the ICL tends to be less prone to discriminate overlapping groups, essentially becoming
an efficient model-based criterion that can be used to outline the clustering structure in the data.

In this paper, we show that in a finite Gaussian mixture model the ICL criterion can be employed using an exact value, rather than the more common approximate variant
exposed in \eqref{ICLdef1} and advocated by \textcite{biernacki2000assessing}.
This is possible thanks to the use of conjugate prior distributions, which essentially allow model parameters to be integrated out apart from the allocation variables. 
Criteria based on exact values of the ICL have already been considered in various contexts: latent class models for categorical data \parencite{biernacki2010exact},
Stochastic Block Models \parencite{come2013model} and Latent Block models \parencite{wyse2014inferring} for network analysis.
Essentially, we show here that a similar framework can be set up for a general finite mixture model, pivoting model choice on an exact formula for the ICL.
We then propose a heuristic greedy algorithm which extends the ones of \textcite{wyse2014inferring,come2013model} and is capable of finding the 
allocations that maximise $f\left( \textbf{x},\textbf{z} \middle\vert K\right)$, hence returning the optimal clustering solution according to the ICL criterion.
Such algorithm has a very little computational cost, making the analysis of very large datasets feasible.

An important advantage of our approach is that it gives a direct answer to the problem of choosing the number of groups $K$, since such value can be inferred straightforwardly from
the optimal allocation vector. 

The paper is organised as follows. In Section \ref{sec:MixtureModels} the modelling assumptions for the mixture models are described. The model is a special case of 
the one used in \textcite{nobile2007bayesian}, and both the univariate and multivariate situations are explored. Section \ref{sec:ExactICL} shows the closed 
form equation for $f\left( \textbf{x},\textbf{z} \middle\vert K\right)$, setting up the framework for the optimization routine. In Section \ref{sec:OptimisingTheExactICL}, the heuristic greedy 
procedure is described in detail: an analysis of the performances and some relevant drawbacks are exposed. 
Section \ref{sec:ChoosingTheHyperparameters} addresses the issue of specifying hyperparameters, while the following two Sections \ref{sec:SimulatedDatasets} 
and \ref{sec:RealDataApplications} provide an application of the algorithm. The paper ends with some final remarks outlined in 
Section \ref{sec:FinalRemarks}.

\section{Mixture models}\label{sec:MixtureModels}
Let $\textbf{x}=\left\{\textbf{x}_1,\dots,\textbf{x}_n \right\}$ be the matrix of observations.
To ease the notation, we describe the case where $\forall\ i=1,2,\dots,n,\ \textbf{x}_i$ is a continuous variable of $b\geq 2$ dimensions, although a 
similar framework can be set up for the univariate case (namely, $b=1$).
In this paper, we assume that the data are realised from a mixture of $K$ multivariate Gaussians, although our approach has more general applicability, and different distributional
assumptions can be considered. 
Hence, the sequence of variables $textbf{x}$ is IID and each component has marginal density:
\begin{equation}
f\left(\textbf{x}_i\middle\vert \boldsymbol{\lambda},\textbf{m},\textbf{R},K\right) = 
\sum_{g=1}^{K} \lambda_g f_b\left( \textbf{x}_i \middle\vert \textbf{m}_{g},\textbf{R}^{-1}_{g}\right)
\end{equation} 
where $f_b\left(\ \cdot\ \middle\vert \textbf{m}_{g},\textbf{R}^{-1}_{g}\right)$ is the multivariate Gaussian density with centre $\textbf{m}_{g}$ and
covariance matrix $\textbf{R}^{-1}_{g}$, for every $g\in\left\{1,\dots,K \right\}$, while $\boldsymbol{\lambda}=\left\{ \lambda_1,\dots,\lambda_K\right\}$ are the mixture weights.
Since we are focusing on a clustering context, it is useful to denote the allocation $g$ of an arbitrary observation $i$ through the variable $z_i=g$, 
so that we have the following expression for the so called complete log-likelihood:
\begin{equation}
\log\left[f\left(\textbf{x}_i,\textbf{z}\middle\vert \textbf{m},\textbf{R},K\right)\right] = 
\sum_{g=1}^{K}\sum_{i:z_i=g} \log\left[\lambda_gf_b\left( \textbf{x}_i \middle\vert \textbf{m}_{g},\textbf{R}^{-1}_{g}\right)\right].
\end{equation} 
We now specify a Bayesian framework as in \textcite{nobile2007bayesian}, setting up a hierarchical structure that defines prior distributions for all 
the parameters involved in the mixture. We assume that the weights $\boldsymbol{\lambda}$ are distributed as a $K$-dimensional Dirichlet variable with hyperparameters 
$\boldsymbol{\alpha}=\left( \alpha_1,\dots,\alpha_K \right)$. 
The allocation variables are IID categorical variables such that for every $i$: 
\begin{equation}
z_i=
 \begin{cases}
  1&\mbox{ with probability }\lambda_1\\
  2&\mbox{ with probability }\lambda_2\\
  \vdots & \hspace{1cm}\vdots\\
  K&\mbox{ with probability }\lambda_K.\\
 \end{cases}
\end{equation}
The remaining parameters are the mixture centres and precisions, which satisfy:
\begin{equation}
\left. \textbf{R}_g \middle\vert \nu, \boldsymbol{\xi}\right. \sim Wishart\left( \nu, \boldsymbol{\xi}\right)
\end{equation}
\begin{equation}\label{GroupMeans}
 \left. \textbf{\textbf{m}}_g \middle\vert \boldsymbol{\mu}, \tau, \textbf{R}_g \right. \sim MVN_b\left( \boldsymbol{\mu}, \left[\tau\textbf{R}_g\right]^{-1}\right)
\end{equation}
So the set of hyperparameters for the model is $\left( \alpha, \tau, \boldsymbol{\mu}, \nu, \boldsymbol{\xi} \right)$, where 
$\alpha$ and $\tau$ are positive real numbers, $\boldsymbol{\xi}$ is a $b\times b$ 
positive definite scale matrix, $\nu>b-1$ are the degrees of freedom, and $\boldsymbol{\mu}$ is a $b$-dimensional vector.

The model described differs from that presented in \textcite{nobile2007bayesian} in that only the symmetric case is considered for the hyperparameters, 
meaning that these do not depend on the group label $g$. The reason for this restriction will be more clear in Section \ref{sec:ExactICL}.

The univariate case has the same structure with a $Ga\left( \gamma,\delta \right)$ distribution replacing the Wishart's, where $\gamma$ and $\delta$
are positive real numbers, and all the parameters are scaled to the proper number of dimensions.

\section{Exact ICL}\label{sec:ExactICL}
The modelling assumptions outlined resemble the very general framework for Gaussian mixture models, with the exception of \eqref{GroupMeans}, 
which states that the 
position of the centre of a component is assumed to be distributed according to the covariance matrix of the component itself. Therefore, groups with similar 
shapes will have similar positioning over the space. Although restrictive, such hypothesis is crucial to allow the collapsing of the model parameters,
i.e., under such assumption, all the mixture's parameters can be integrated out, returning the complete log model evidence in an exact form.
Nonetheless, such approach has already been used in other important works, such as \textcite{nobile2007bayesian} and \textcite{steele2010performance}.
More in detail, in \textcite{nobile2007bayesian}, such feature is exploited to set up a collapsed Gibbs sampler (denoted \textit{allocation sampler}) 
that draws from the posterior distribution of the allocations and number of groups.

Here, we can obtain the same factorization of the complete log model evidence:
\begin{equation}\label{ICLeq1}
 \begin{split}
   f\left(\textbf{x},\textbf{z}\middle\vert \boldsymbol{\alpha}, \boldsymbol{\phi}, K\right) &=
   \int \int \int f\left(\textbf{x},\textbf{z}, \textbf{m},\textbf{R}, \boldsymbol{\lambda}\middle\vert \boldsymbol{\alpha}, \boldsymbol{\phi}, K\right)d\textbf{m}d\textbf{R}d\boldsymbol{\lambda}\\
  &=  \int \int \int f\left(\textbf{x}\middle\vert \textbf{z}, \textbf{m},\textbf{R},K\right)\pi\left(\textbf{z}\middle\vert \boldsymbol{\lambda},K\right)
  \pi\left(\boldsymbol{\lambda}\middle\vert \boldsymbol{\alpha},K\right)\pi\left(\textbf{m},\textbf{R}\middle\vert \boldsymbol{\phi},K\right)d\textbf{m}d\textbf{R}d\boldsymbol{\lambda}\\
  &=  \int\int f\left(\textbf{x}\middle\vert \textbf{z}, \textbf{m},\textbf{R},K\right)\pi\left(\textbf{m},\textbf{R}\middle\vert \boldsymbol{\phi},K\right)d\textbf{m}d\textbf{R}
   \int\pi\left(\textbf{z}\middle\vert \boldsymbol{\lambda},K\right)\pi\left(\boldsymbol{\lambda}\middle\vert \boldsymbol{\alpha}, K\right)d\boldsymbol{\lambda}\\
  &=   f\left(\textbf{x}\middle\vert \textbf{z},\boldsymbol{\phi}, K\right)
   \pi\left(\textbf{z}\middle\vert \boldsymbol{\alpha},K\right),
 \end{split}
\end{equation}
where $\boldsymbol{\phi}=\left( \tau, \boldsymbol{\mu}, \nu, \boldsymbol{\xi} \right)$ are the hyperparameters for the model.
Then we take advantage of the specific model assumptions to collapse the model parameters and hence obtain exact formulas for the logs of the final terms on the right hand side of \eqref{ICLeq1}:
\begin{equation}\label{ICLeq3}
 \begin{split}
  \log f\left(\textbf{x}\middle\vert \textbf{z},\boldsymbol{\phi}, K\right)&= 
  \sum_{g=1}^{K} \Bigg\{-\frac{bn_g}{2}\log(\pi) + \frac{b}{2}\log(\tau) - \frac{b}{2}\log(\tau+n_g)\\
  &+\sum_{s=1}^{b} \left[\log\Gamma\left( \frac{\nu+n_g+1-s}{2} \right) - \log\Gamma\left( \frac{\nu+1-s}{2} \right)\right]
  +\frac{\nu}{2}\log\left|\boldsymbol{\xi}\right| \\
  &-\frac{\left( \nu+n_g \right)}{2}\log\left|\boldsymbol{\xi}+\sum_{i:z_i=g}(\textbf{x}_i-\bar{\textbf{x}}_g)(\textbf{x}_i-\bar{\textbf{x}}_g)^t
  +\frac{\tau n_g}{\tau+n_g}(\bar{\textbf{x}}_g-\boldsymbol{\mu})(\bar{\textbf{x}}_g-\boldsymbol{\mu})^t\right|
  \Bigg\}
 \end{split}
\end{equation}
\begin{equation}\label{ICLeq}
 \begin{split}
  \log \pi\left(\textbf{z}\middle\vert \boldsymbol{\alpha},K\right)&=
 \log\Gamma(K\alpha) - \log\Gamma(K\alpha+n) - K\log\Gamma(\alpha) + \sum_{g=1}^{K}\log\Gamma(\alpha+n_g)
 \end{split}
\end{equation}
where $n_g$ is the number of observations and $\bar{\textbf{x}}_g$ is the centre of group $g$, while $|\textbf{A}|$ denotes the determinant of an arbitrary matrix $\textbf{A}$.
The exact value of the ICL is thus given by:
\begin{equation}\label{exactICLdef1}
 \begin{split}
   ICL_{ex}(\textbf{z},K) &=
  \log f\left(\textbf{x}\middle\vert \textbf{z},\boldsymbol{\phi}, K\right)  + 
   \log\pi\left(\textbf{z}\middle\vert \boldsymbol{\alpha},K\right).
 \end{split}
\end{equation}

Clearly the $ICL_{ex}$ depends on the hyperparameters $\alpha$ and $\boldsymbol{\phi}$. This does not happen when the same value is approximated 
using BIC, since in such case the model parameters are replaced with Maximum Likelihood Estimates that, by definition, depend only on the data. 
Hence, every result obtained through $ICL_{ex}$ will depend on the hyperparameters, which must be chosen wisely and a priori.

Once hyperparameters are defined, the $ICL_{ex}$ will simply be a function of the data (which is fixed) and of the allocations.
Therefore, it is possible to search over the space of all the possible clustering configurations to find one maximizing the $ICL_{ex}$ value.
Such partition can be regarded as an appropriate clustering solution since it naturally maximises a model-based clustering criterion, taking into account both
the modelling assumptions and the degree of separation among groups.
Furthermore, the ICL criterion is completely fulfilled in that the corresponding exact value (rather than an approximation) is maximised, hence 
yielding an obvious advantage in terms of reliability.

A key advantage of the framework exposed is that the $ICL_{ex}$ function depends on $K$ through the allocation variables. 
Indeed, during the optimization, we can exploit the fact that the allocations $\textbf{z}=\left\{z_1,\dots,z_n\right\}$ are categorical variables to
recover exactly the value of $K$, which will be equal to the number of different states of $\textbf{z}$.
However, this convenience comes to a price, since all the hyperparameters depend on $K$ as well. 
Since these are supposed to be fixed apriori, a conflict arises between the number of dimensions of the hyperparameters and the value of $K$ inferred 
from the allocations at each step of the optimization process. 
To avoid these issues all at once, we simply restrict the finite mixture model using the aforementioned symmetric hypothesis, pretending that the 
hyperparameters corresponding to each model have the same value on every dimension.
Of course, other approaches can be used to overcome the same problem. As an example, one could simply specify a possibly asymmetric set of hyperparameters for each 
value of $K$. Similarly to the finite mixture distributional assumptions, we do not explore this possibility in this work for the sake of brevity, and leave the task
to possible future works.

Once optimization has been performed, the resulting configuration $\hat{\textbf{z}}$ maximises the $ICL_{ex}$ and thus is selected 
as the clustering solution. 
Note that, although we use a parametric framework, the mixture parameters are not included in the solution. 
Indeed, the algorithm we describe in this paper is meant to serve mainly as a direct tool to obtain a clustering of the data
and an estimation of the number of clusters, taking advantage of the fact that it has the potential to scale well to very large datasets.
Also, in contrast to the standard model-based clustering for Gaussian mixtures \parencite{fraley2002model,biernacki2000assessing}, no constraints on the covariance matrices of the components are considered.

At this stage, two main points must be addressed. The first is whether we are always able to find the global optimum of the $ICL_{ex}$ function,
and, in case, how do we find it. The second problem concerns the choice of hyperparameters, and how this decision affects the final
solution. Both issues will be tackled in the following Sections.

\section{Optimising the exact ICL}\label{sec:OptimisingTheExactICL}
The number of possible clustering configurations in the search space for the optimization routine is of the order of $O((K_{max})^n)$, 
where $K_{max}$ is the maximum number of groups allowed. A complete exploration of such space is not possible, even for small $K_{max}$ and $n$. 
However, similar combinatorial problems have been already faced in other works, such as \textcite{newman2004fast},
by means of heuristic greedy routines that resemble the well known Iterated Conditional Modes of \textcite{besag1986statistical}. 
We mention also a different approach relying on genetic algorithms, proposed in \cite{tessier2006evolutionary} to find the clustering configuration maximising an exact ICL value in a latent class model.
More recently, greedy routines have been employed in \textcite{come2013model} and \textcite{wyse2014inferring} to maximise $ICL_{ex}$ with respect to the 
allocations on Stochastic Block Models and Bipartite Block Models for networks, respectively. 
In both these works the number of groups is inferred from the allocations. 
We now give a description of the main idea of the greedy optimization routine, applying it to our Gaussian mixture clustering problem.

The first step consists of finding a configuration to initialise the algorithm. Such initial solution usually has $K_{max}$ number of groups, and can
be chosen arbitrarily. Random allocations are acceptable, however more informative clustering solutions can be provided to ensure correct 
convergence. In this work, as starting configuration, we use random allocations exclusively.
Then, essentially, the algorithm iterates over the observations in a random order and updates each categorical allocation choosing the value that 
gives the best update according to the objective function. Once a complete loop over all the observations does not
yield any increase in the $ICL_{ex}$ the algorithm is stopped. Pseudocode is provided in Algorithm \ref{GreedyICL}. In the pseudocode we denote
by $\textbf{z}_{A\rightarrow g}$ the configuration $\textbf{z}$ where the allocations corresponding to the indexes in the set $A$ are changed into $g$, and 
$\ell_{A\rightarrow g}$ is the corresponding updated $ICL_{ex}$.

\begin{algorithm}[htb]
\begin{spacing}{1.2}
\begin{algorithmic}
\State Initialise $\textbf{z}$ 
\For{every iteration}
\State Let $\textbf{z}$, $K$, $\ell$ be the current allocations, number of groups and $ICL_{ex}$, respectively.
\State Let $\ell_{stop}=\ell$
\State Initialise $V=\left\{ 1,2,\dots,n\right\}$ as the pool of the observation labels
\While{$V$ not empty}
\State Pick randomly an observation $i$ from $V$ and delete it from $V$
\State $\hat{g}=\displaystyle\argmax_{\substack{g=1,2,\dots,K+1}} \ell_{i\rightarrow g}$
\State $\ell=\ell_{i\rightarrow \hat{g}}$
\State $\textbf{z}=\textbf{z}_{i\rightarrow \hat{g}}$
\State Reassign the labels of $\textbf{z}$ so that the smallest numbers possible are used
\State $K = \max\textbf{z}$
\EndWhile
\If {$\ell_{stop}=\ell$} \State \emph{break} \EndIf
\EndFor
\State Return $\textbf{z}$, $K$ and $\ell$
\end{algorithmic}
\caption{Greedy algorithm}
\label{GreedyICL}
\end{spacing}
\end{algorithm}

\begin{figure}[htb]
\centering
 \includegraphics[width=0.24\textwidth]{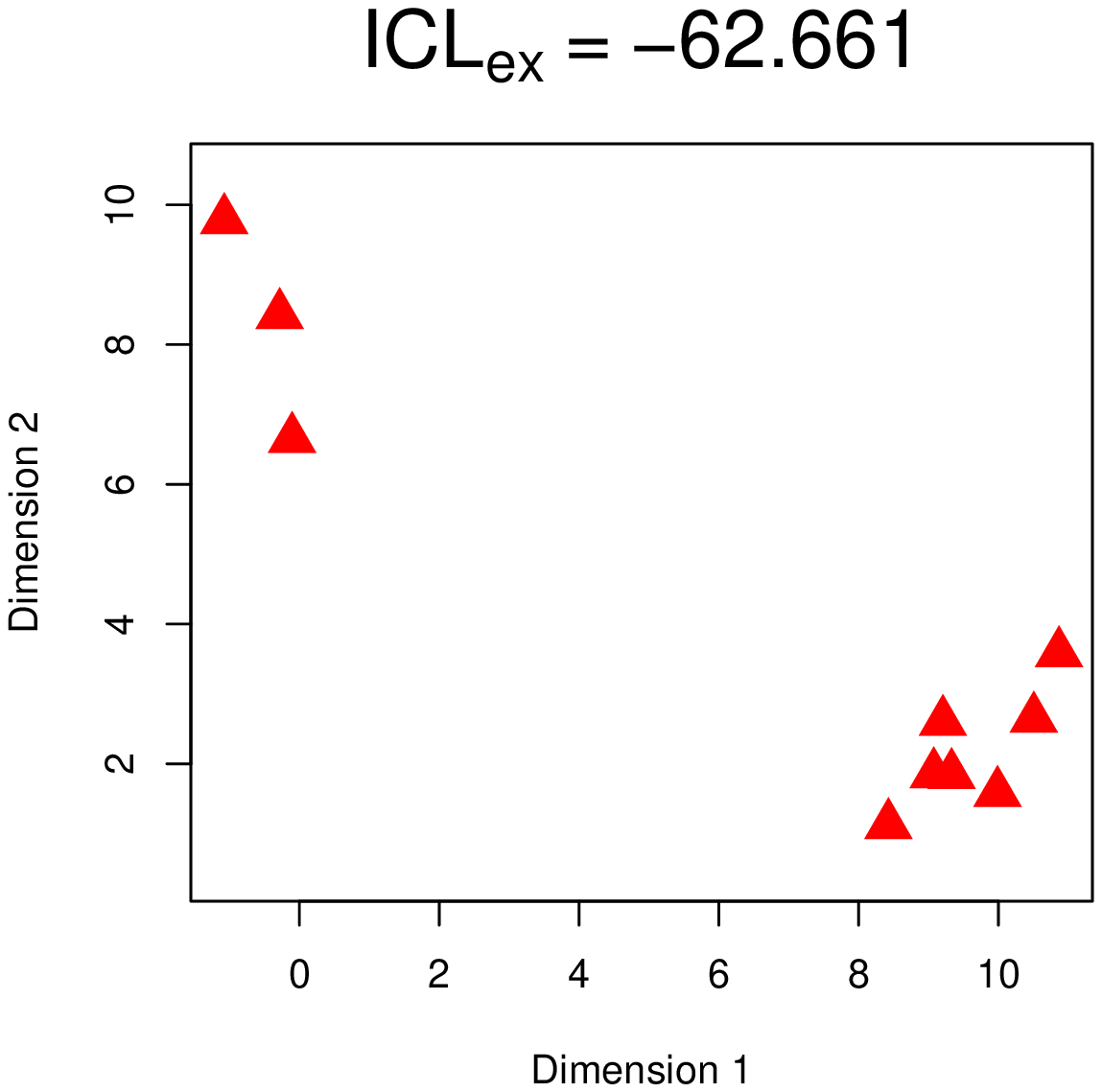}
 \includegraphics[width=0.24\textwidth]{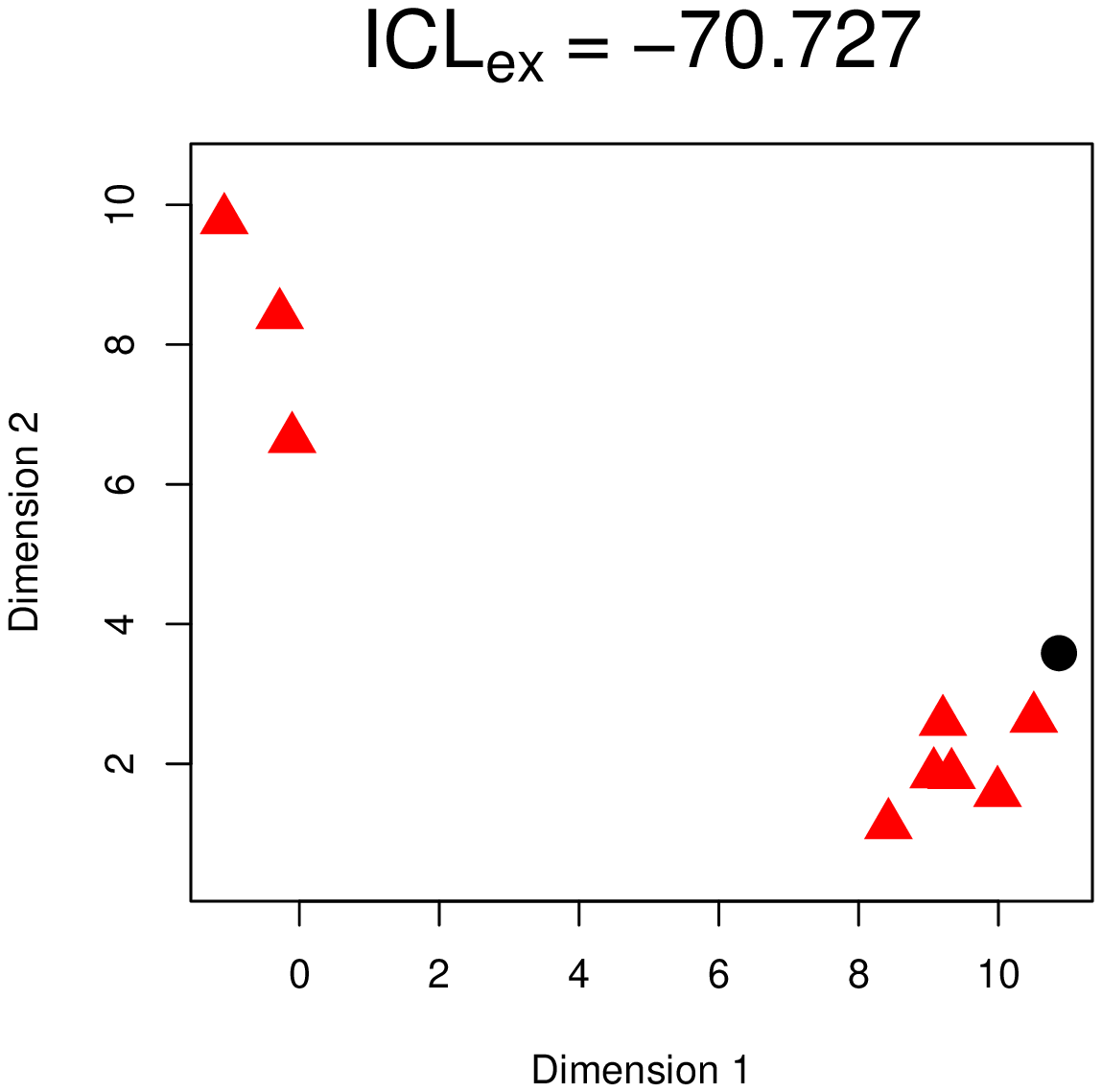}
 \includegraphics[width=0.24\textwidth]{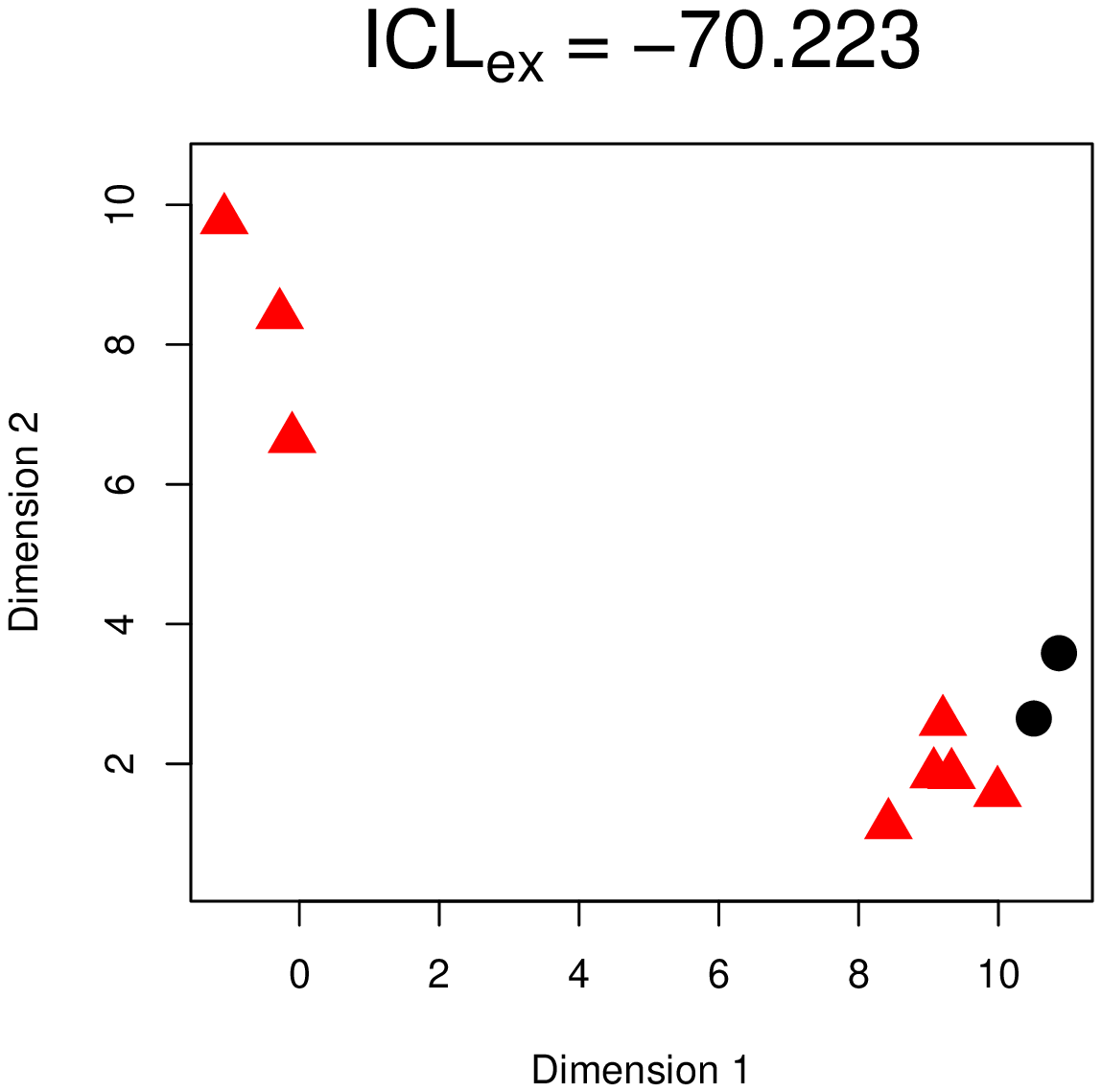}
 \includegraphics[width=0.24\textwidth]{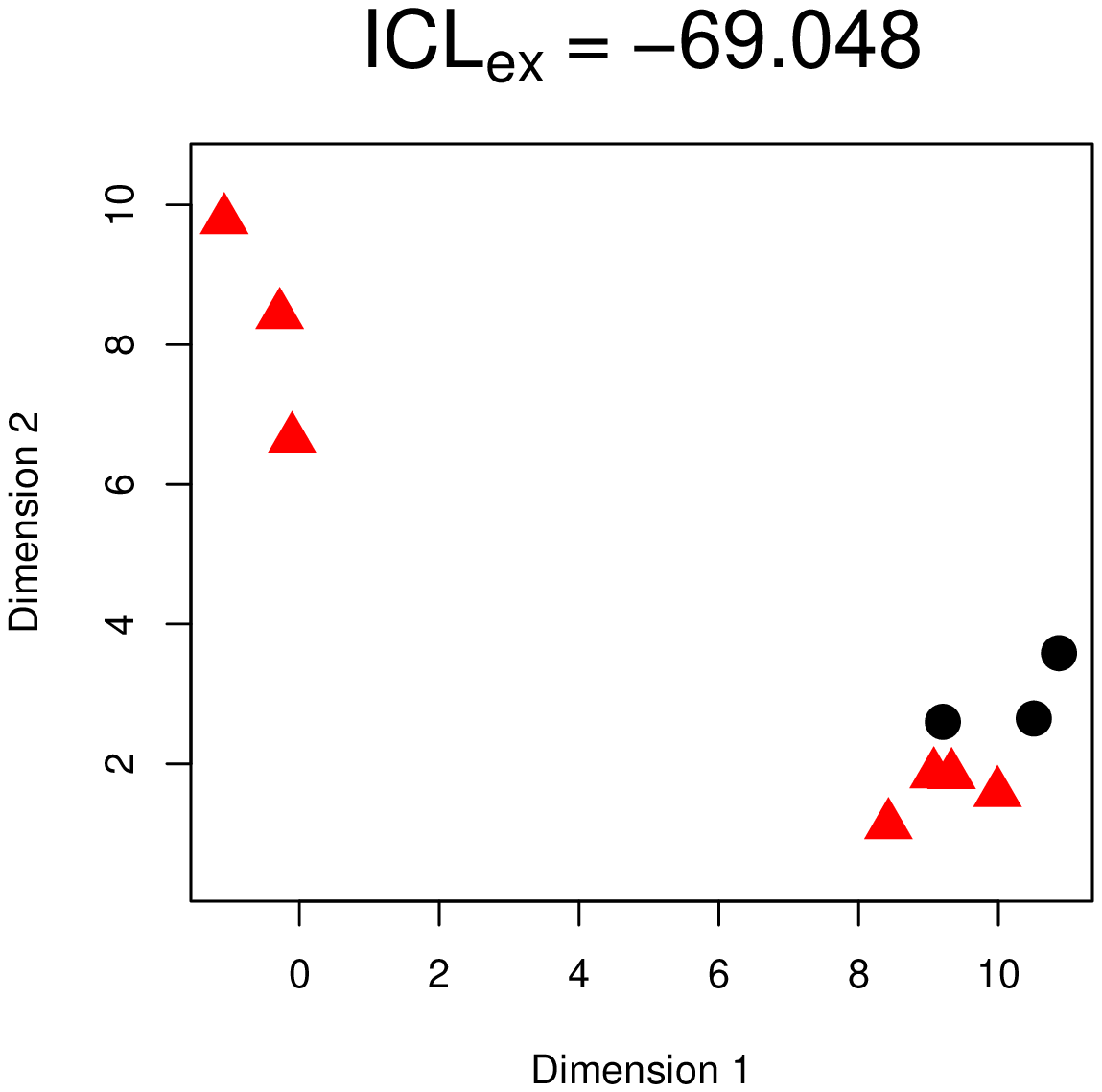}
 \includegraphics[width=0.24\textwidth]{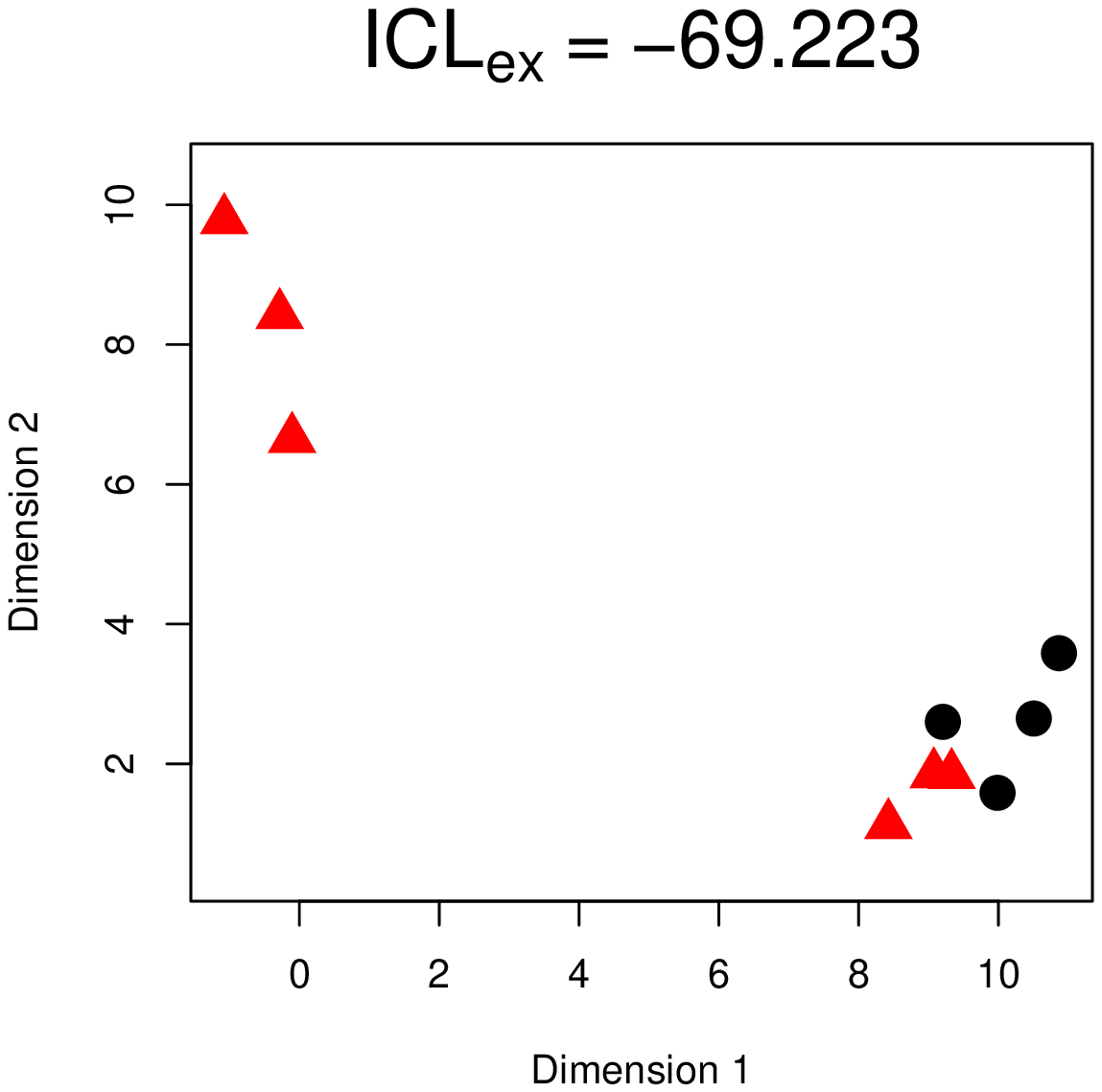}
 \includegraphics[width=0.24\textwidth]{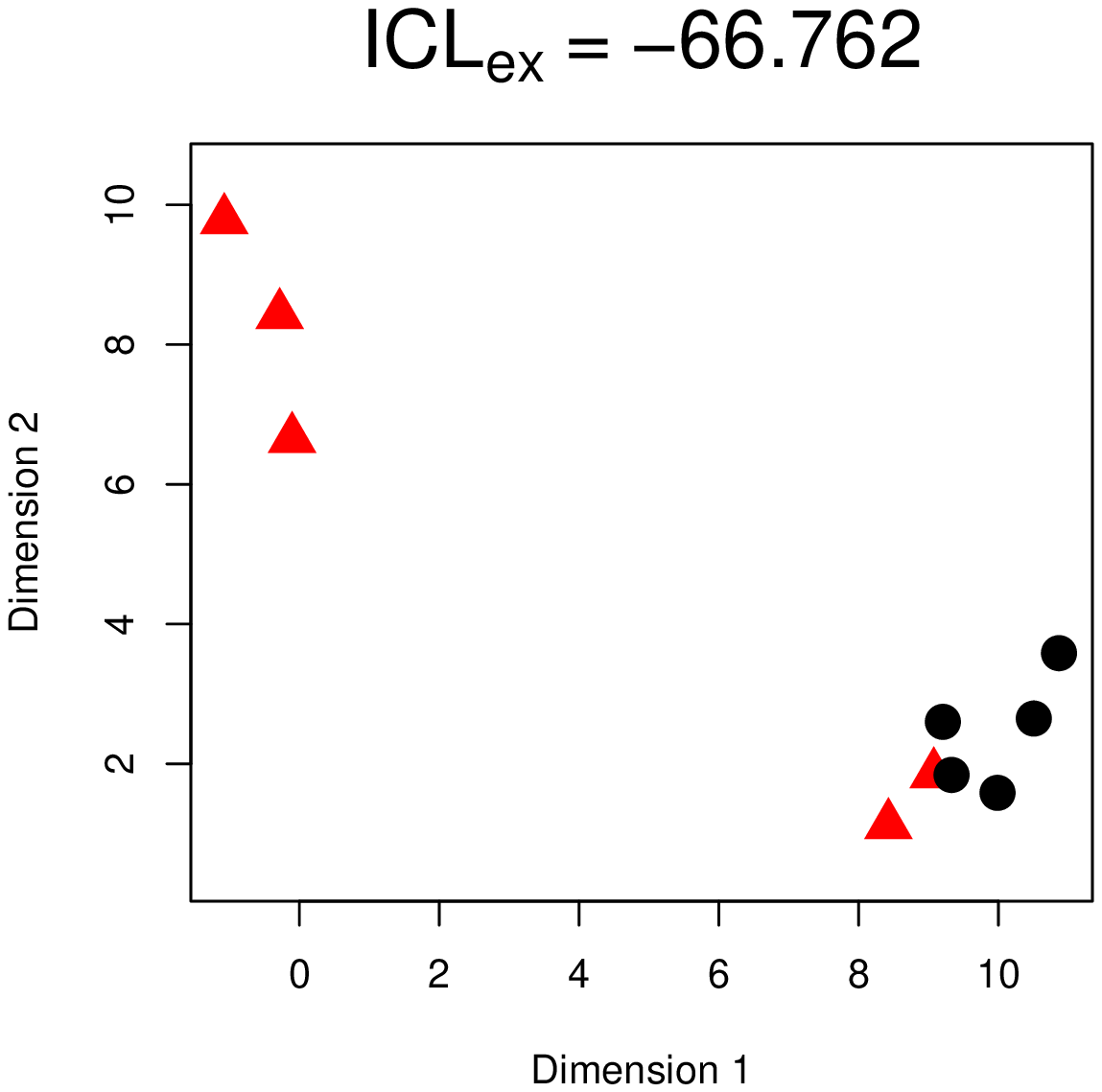}
 \includegraphics[width=0.24\textwidth]{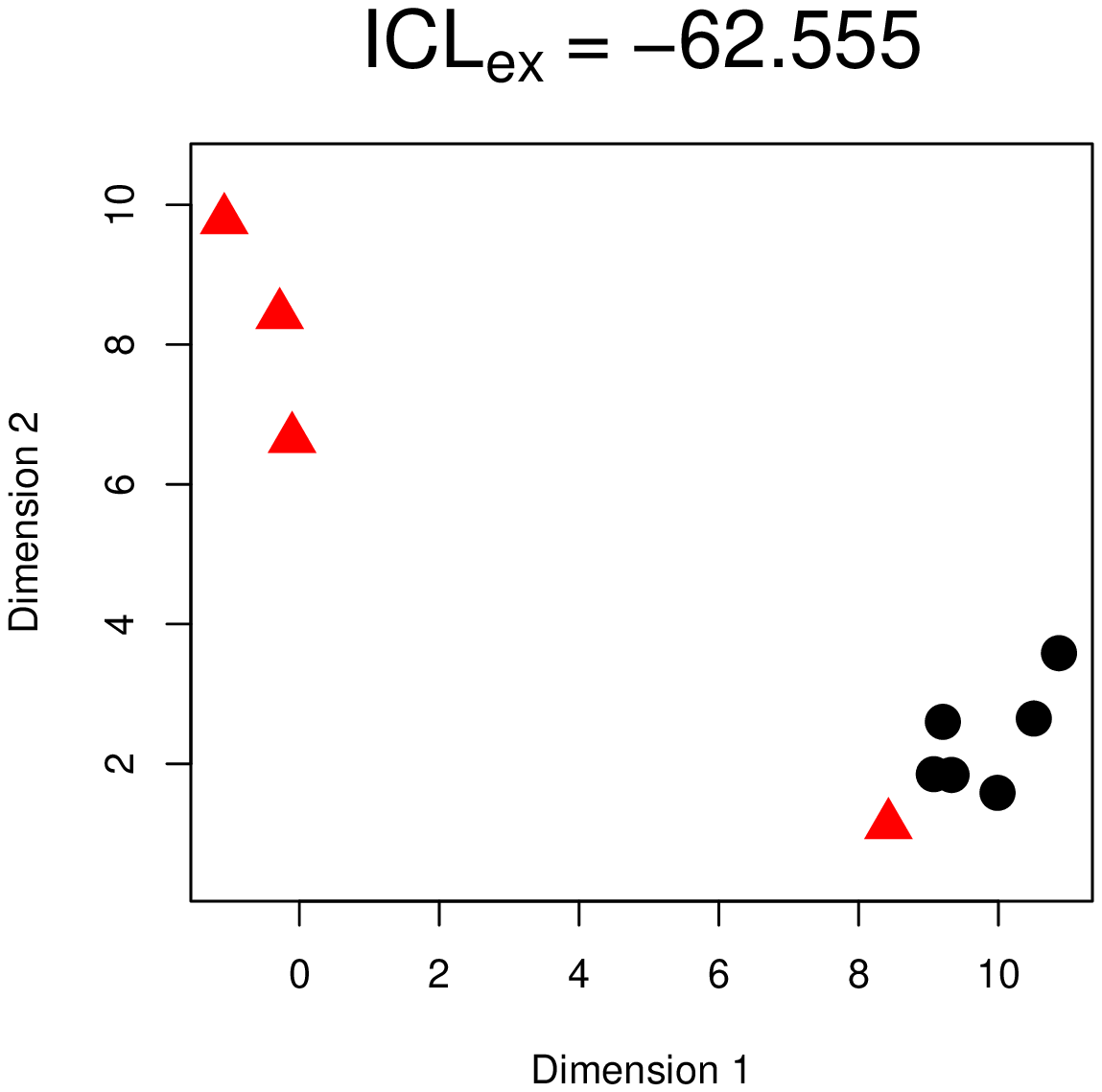}
 \includegraphics[width=0.24\textwidth]{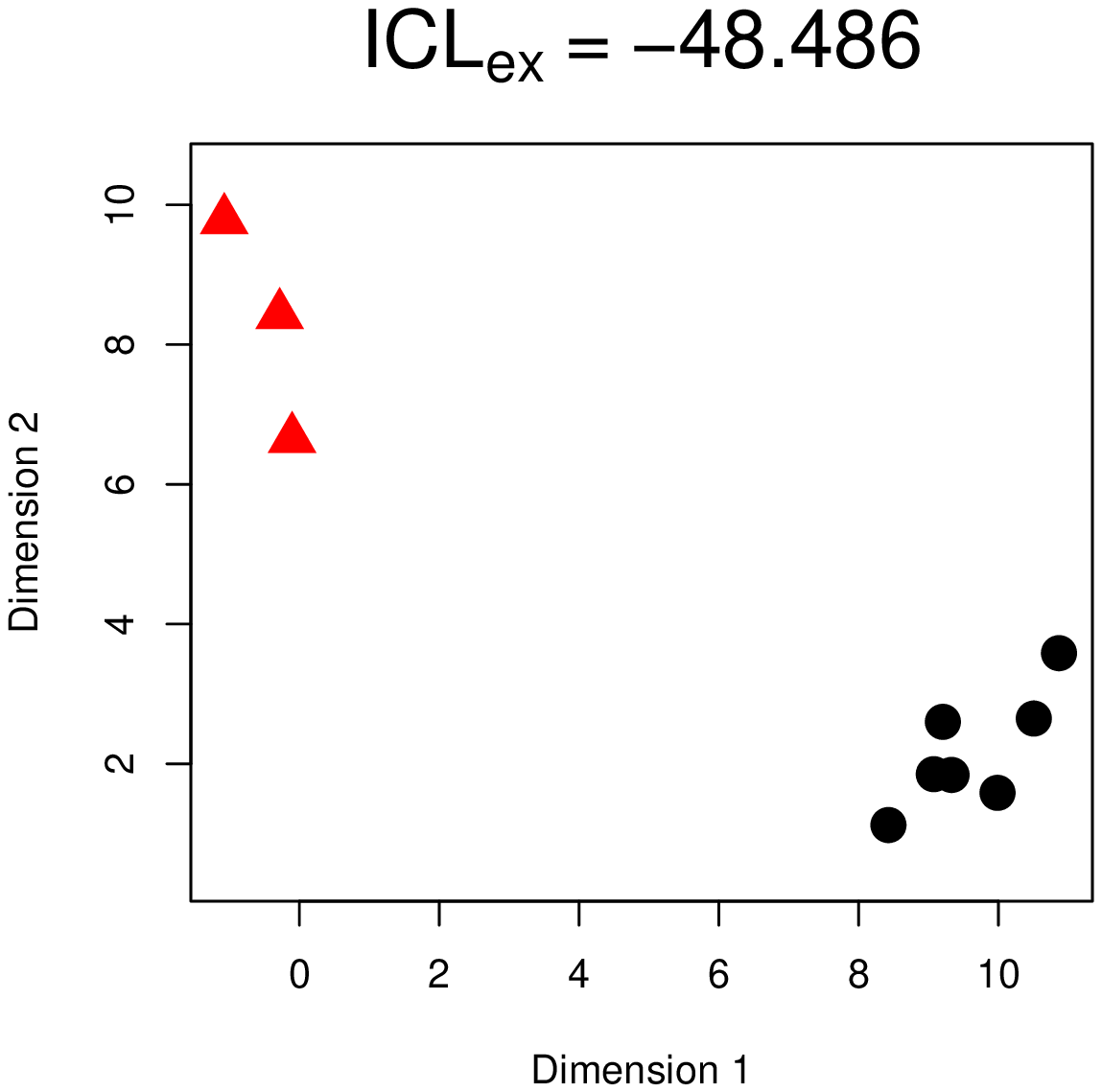}
 \caption{$ICL_{ex}$ is very sensitive to local optima: in this dataset, composed of $10$ observations only, in order to have an increase in the objective 
 function and thus being able to leave the first configuration on the top left, at least $6$ observations must be reallocated at the same time. 
 The values chosen as hyperparameters are the same ones used to generate the data.}
 \label{fig:CombinedStep}
\end{figure}

The algorithm described (denoted from now on \textbf{Greedy ICL}), 
dramatically reduces the search space, usually reaching convergence in very few steps. 
A key feature is that it works very well in merging clusters, i. e. during the very first iteration most of the groups present are emptied completely.
Hence, at the end of the first loop over the observations, the current configuration is made of a very small number of groups.
As soon as most of the groups have been collapsed and only few large clusters remain, the algorithm tends to get stuck due to the poor
sensitivity of $ICL_{ex}$ with respect to little changes. Actually, such a configuration might very likely be a local optimum, rather than the global one.
In other words, assuming that a better solution is obtainable by splitting or merging groups, the Greedy ICL will 
not be able to discover it, mainly because the objective function does not increase when only one observation at a time is reallocated 
(groups of small sizes make an exception, for obvious reasons). 
Figure \ref{fig:CombinedStep} is intended to explain this behaviour by a trivial example.
To tackle this issue, in \textcite{come2013model} the authors propose multiple reruns of the routine with different starting configurations
and a final merge step with the explicit purpose of avoiding local optima. 
However, this final merge step does not take into account the fact that a better solution could also be obtained through a splitting of the current groups,
making an underestimation of $K$ rather frequent (at least in the Gaussian finite mixture context).

With regard to the computational cost, an exact quantification is not possible since the number of iterations needed for convergence is random. 
However, iteration-wise we can state that $O(nK_{max})$ evaluations of the objective function are needed.
Also, for datasets containing less than $5$ groups and $100$ observations typically less than $5$ iterations grant proper convergence. 
The final merging has a cost of $O(\hat{K}^3)$ where $\hat{K}$ is the number of groups in the final configuration \parencite{come2013model}.
A very interesting feature of the greedy routine exposed is that the evaluation of the objective function is actually needed only once at the beginning of the algorithm,
since then the $ICL_{ex}$ can be updated calculating only the variation induced by the single reallocation, which has a negligible computational cost.
However, we do not focus here on the computational aspects of the greedy routines.

\subsection{Greedy Combined algorithm}\label{sec:GreedyCombinedICL}
In this paper we introduce a variation for the Greedy ICL algorithm to improve its performances. Attention will be mainly focused on the 
quality of the final solution, rather than on the computational cost needed. However, the idea used does not make the algorithm noticeably slower.

As shown, the main drawback of the Greedy ICL is its high sensitivity to local optima. This comes as a consequence of the fact that observations
are reallocated one at a time, whereas very often a more radical change in the current configuration is needed to yield an appreciable increase in the 
objective function.
The Greedy ICL algorithm gets indeed stuck very easily and once that happens the main loop breaks. 
In other words, the steps on the search space are not wide enough to allow a proper exploration of the space itself. 
Therefore, we propose an algorithm that exhibits the same greedy behaviour, but makes use of larger steps, allowing a better exploration and (to some degree) 
the ability to leave local optima as well. 
We define these wider steps as combined updates of multiple allocations. 
Thus, our algorithm is essentially the same Greedy ICL algorithm where instead of updating the allocation of one observation at a time, 
a set of allocations is updated. 
The observations in such set must belong to the same group, and are chosen using a nearest neighbour criterion. Furthermore, the set is updated as a block, therefore,
for each iteration, only $O(nK_{max})$ evaluations of the objective function are actually required.
The number of observations chosen is the realization of a Beta-Binomial random variable, where the Beta hyperparameters are user defined 
and the number of trials is given by the cardinality of the group where the observations are currently allocated.
In terms of computational cost, the number of iterations needed tends to be higher. Also, although the number of evaluations of the objective function is 
exactly the same, the trick of reducing the cost by using the updates only will be definitely affected negatively.
From now on, the algorithm just described will be denoted by \textbf{Greedy Combined ICL}. Pseudocode is provided in Algorithm \ref{GreedyCombinedICL}. 

\begin{algorithm}[htb]
\begin{spacing}{1.2}
\begin{algorithmic}
\State Initialise $\textbf{z}$ 
\For{every iteration}
\State Let $\textbf{z}$, $K$, $\ell$ be the current allocations, number of groups and $ICL_{ex}$, respectively
\State Initialise $V=\left\{ 1,2,\dots,n\right\}$ as the pool of the observation labels
\While{$V$ not empty}
\State Pick randomly an observation $i$ from $V$ and delete it from $V$
\State Let $J_i = \left\{j_1,\dots,j_{K_i}\right\}$ be the set of observations allocated in the same group as $i$, ordered increasingly according to their distance 
with respect to $i$
\State Sample $\eta$ from a $Beta(\beta_1,\beta_2)$
\State Sample $r$ from a $Bin\left( K_i, \eta \right)$
\State Let $J'_i = \left\{j_1,\dots,j_r\right\}$ be the set of the first $r$ observations in $J_i$
\State $\hat{g}=\displaystyle\argmax_{\substack{g=1,2,\dots,K+1}} l_{J'_i\rightarrow g}$
\If{$\ell<\ell_{i\hat{g}}$}
\State $\ell=\ell_{i\hat{g}}$
\State $\textbf{z}=\textbf{z}_{i\rightarrow \hat{g}}$
\State Reassign the labels of $\textbf{z}$ so that the smallest numbers possible are used
\State $K = \max\textbf{z}$
\EndIf
\EndWhile
\EndFor
\State Return $\textbf{z}$, $K$ and $\ell$
\end{algorithmic}
\caption{Greedy Combined algorithm}
\label{GreedyCombinedICL}
\end{spacing}
\end{algorithm}


It must be pointed out that in the Greedy Combined algorithm the distance between observations is used to create the set of neighbours $J'_i$. Thus, a matrix of dissimilarities must be 
evaluated at the beginning of the procedure for a storage and computational cost proportional to $O(n^2)$. 
The distance used is chosen by the user. Usually the Euclidean
distance does a fine job, however proper transformations of the data and different distance choices might be advisable when clusters do not have a spherical
shape. Indeed, a more round shaped cluster will definitely couple better with the nearest neighbour concept.

\section{Choosing the hyperparameters}\label{sec:ChoosingTheHyperparameters}
As defined in Section \ref{sec:MixtureModels}, in the multivariate case the hyperparameters to be set are 
$\alpha$, $\tau$, $\boldsymbol{\mu}$, $\nu$ and $\boldsymbol{\xi}$. As already mentioned, the approximate version of
ICL introduced in \textcite{biernacki2000assessing} does not require the specification of any prior distribution, simplifying the interpretation of the maximised clustering 
configuration. 
In our case, instead, the objective function to be maximised does depend on hyperparameters, which subsequently can affect the solution obtained considerably.
In the standard Bayesian approach, hyperparameters should be chosen in the way that best represents the prior information that one possesses.
At the same time, one could be interested in specifying non-informative distributions, since prior information is not always available.
However, as shown in \textcite{jasra2005markov}, this special case is difficult to represent in the standard Gaussian mixture model context. This issue apparently extends
to the model described in this paper as well.
With this in mind, we limit ourselves to describing some guidelines to interpret the hyperparameters and thus to transform the prior information possessed in an optimal way.

Firstly, we recall that a symmetric assumption has been made. This is convenient since it removes the dependence of the hyperparameters on $K$.
Essentially, we are assuming that for every value of $K$, each hyperparameter vector has equal entries.

Now, as concerns the group proportions, these are a realization of a Dirichlet
random vector characterised by the $K$-dimensional parameter vector $\left( \alpha,\dots,\alpha \right)$. 
The value $\alpha=0.5$ corresponds to a non-informative Jeffreys' prior \parencite{jeffreys1946invariant}, whereas the value $4$ has been advocated in other well known
works (\cite{mengersen2011mixtures}, chapter 10). Essentially, a larger value will promote groups of equal sizes, while a smaller value will support the rise of a single large group.
In this paper, we will consider the default value $\alpha=4$.

As concerns the hyperparameter $\boldsymbol{\mu}$, this can be simply chosen as the centre of the observed data.

The hyperparameter $\nu$ is constrained to be greater than $b-1$, and it describes the shape that the clusters are supposed to have. Values of $\nu$ close to 
$b$ will support extremely narrow elliptical shapes whereas higher values will support more rounded shapes. As a default choice we propose $\nu=b+1$, which
can account for many different settings. However, according to prior beliefs, any value ranging from $b$ to $b+10$ can be chosen.

\begin{figure}[htb]
\centering
 \includegraphics[width=0.45\textwidth]{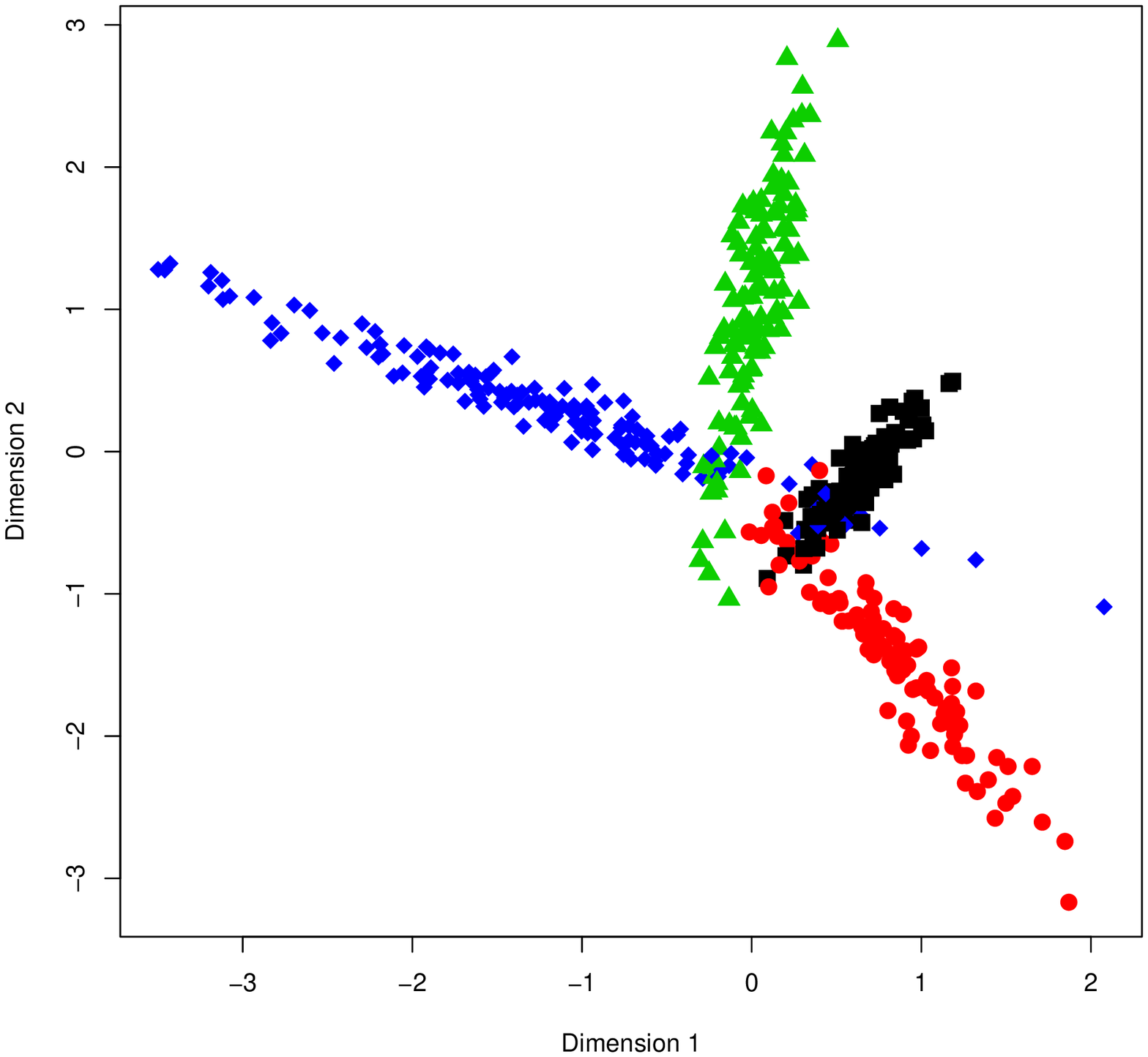}
 \includegraphics[width=0.45\textwidth]{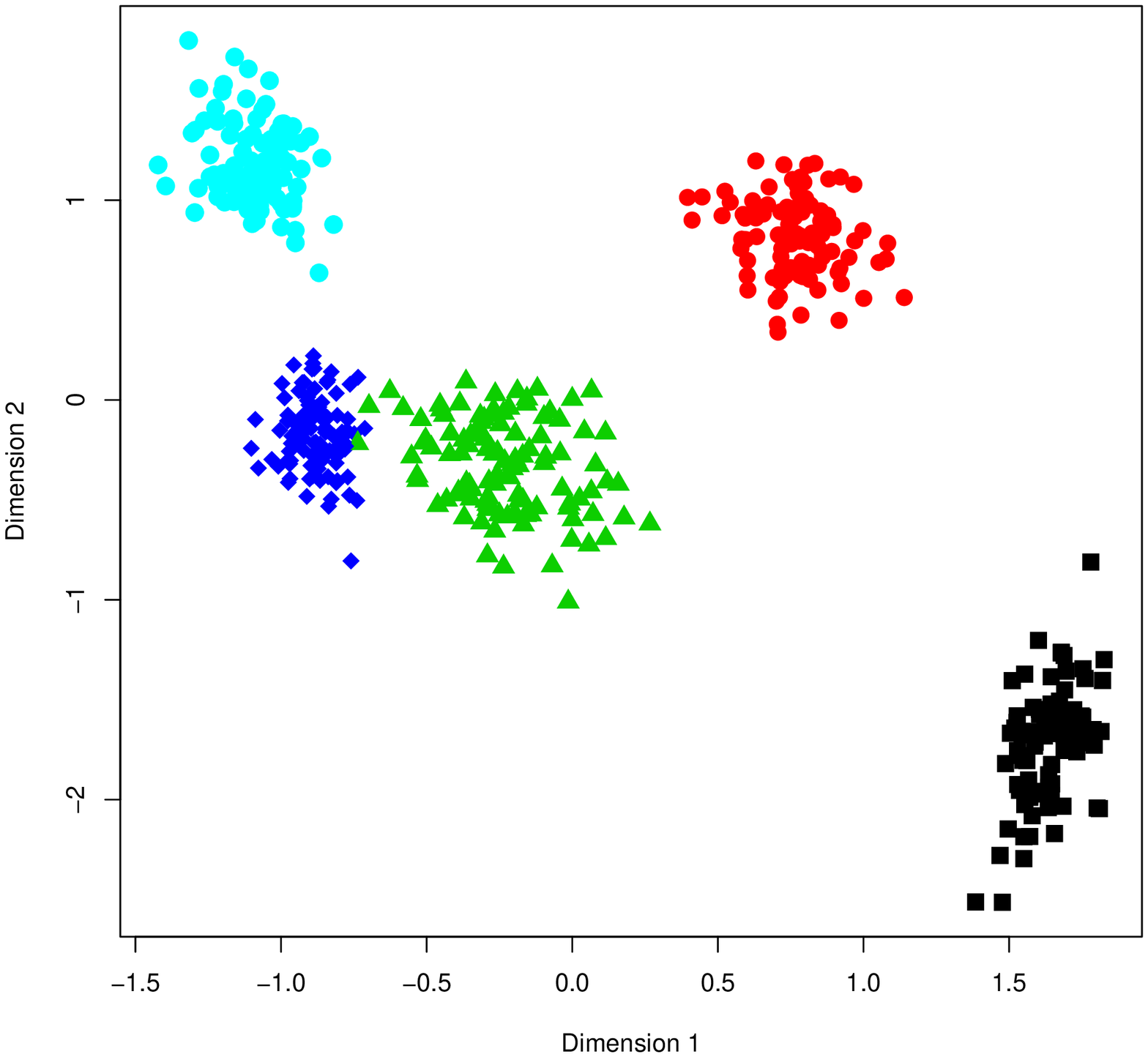}
 \caption{$500$ points divided in $5$ groups and generated according to the model described in Section \ref{sec:MixtureModels}. Common hyperparameters
 are $\alpha=100,\ \boldsymbol{\mu}=(0,0)^t$ and the identity matrix as $\boldsymbol{\xi}$. On the left hand image $\tau=0.1$ and $\nu=b=2$ while on the right
 hand image $\tau=0.01$ and $\nu=b+10$. $\omega$ is omitted from the example since it has no effect when data are standardised.} 
 \label{fig:HyperEx}
\end{figure}

The remaining hyperparameters $\tau$ and $\boldsymbol{\xi}$ are closely related, thus more difficult to choose.
We make a diagonal assumption on $\boldsymbol{\xi}$, which does not affect particularly the range of situations described. 
One must choose, then, the parameter $\tau$ and the diagonal element which will be here denoted $\omega$. 
While $\tau$ determines how separated the clusters' centres are, $\omega$ describes the volumes of the covariance matrices for 
both the data points and the cluster centres. In other words, the covariance matrix of $\textbf{x}$ is affected only by $\omega$, while the covariance matrices
for the clusters centres' positions are determined by both $\tau$ and $\omega$. 
Then, different values of these two hyperparameters will yield different combinations of clusters' sizes and clusters' separation.
Possible values include $\tau=0.1,\ \tau=0.01,\ \tau=0.001,\ \omega=0.1,\ \omega=1,\ \omega=10$.
Figure \ref{fig:HyperEx} shows two generated datasets corresponding to particular choices of $\tau$ and $\nu$, which yield datasets
with different cluster shapes and different degrees of separation.\\
For the univariate case, we propose as default hyperparameters $\gamma=\delta=0.5$, which can account for a wide range of cases, however,
different ratios $\gamma/\delta$ values will have a meaning equivalent to that of $\omega$.

A final note concerns the choice of the parameters for the Beta-Binomial distribution. We propose the values $\beta_1=0.1$ and $\beta_2=0.01$ as default
choices, for datasets made of a few hundreds observations like the ones shown in this paper. These values are completely up to the user, and for most applications
they do not need careful tuning: default values will work just fine. 
However, when groups are very concentrated and overlapping it may be advisable 
to tune up the distribution of the number of neighbours to speed up the routine and ensure better convergence. 
In particular, the best results are obtained when 
the number of neighbours updated is comparable to the size of the group that should be recognised. 

\section{Simulated datasets}\label{sec:SimulatedDatasets}
We propose in this Section an application of the methodology described to several simulated datasets.
Various comparisons with other clustering methods are considered. For more clarity, we use throughout the following notation:
\begin{itemize}
 \item \textbf{GCICL} denotes the solution optimising $ICL_{ex}$ obtained using the Greedy Combined ICL algorithm.
 \item \textbf{MBCBIC} denotes the optimal configuration where the model parameters are maximum likelihood estimates obtained through the Expectation-Maximization 
 algorithm, and the model is chosen through the BIC criterion \parencite{fraley2002model}.
 \item \textbf{MBCICL} denotes the optimal configuration where the model parameters are maximum likelihood estimates obtained through the Expectation-Maximization 
 algorithm, and the model is chosen through the approximate version of the ICL of \parencite{biernacki2000assessing}.
\end{itemize}
The purpose of these comparisons is not meant to create a ranking between the different approaches. Indeed the various methods work differently and have different 
modelling assumptions: the two model-based clustering procedures impose a factorization of the covariance matrix of the components that allows the use of simpler models.
Also, the maximization of $ICL_{ex}$ has a more subjective nature, since it depends on hyperparameters' choices.
We would like to make one point clear: in the simulated studies, the data used is generated using the likelihood model described in Section \ref{sec:MixtureModels}. The MBC
routines (parameters are MLE obtained through the Expectation Maximization algorithm) perform model choice using the BIC, which is supposed to return the best model according to the 
information contained in the likelihood.
Thus, since the data is generated from the likelihood model itself, it makes sense to compare the MBC solutions with the actual ``true'' allocations
generated, to assess the methods in terms of efficiency. On the other hand, the maximization of the exact ICL has a completely different nature, since here we are maximizing 
(the log of) the quantity $p\left( \textbf{z},\textbf{x}\middle\vert K \right)$
which, as a function of $\textbf{z}$, is proportional to the posterior distribution of the allocations. 
Thus maximizing $ICL_{ex}$ corresponds to obtaining the MAP values for the allocations.
As a consequence, the global optimum of $ICL_{ex}$ is not maximizing a likelihood, but rather the prior distribution updated through the likelihood, i.e. a posterior distribution. 
Hence, comparing the configuration maximizing $ICL_{ex}$ to the MBC ones or the true generated one has little meaning, since we are not maximizing the same quantity.
It must be noted, however, that under the choice of non-informative prior distributions, the MAP configuration is expected to be determined exclusively by the information provided by the likelihood,
making the method more objective and suitable for comparisons in terms of performances. However, as pointed out in previous important contributions (see for example \cite{jasra2005markov}),
non-informative prior distributions are not available for the model considered. This is in contrast to other works based on the maximization of the exact version of ICL 
\parencite{biernacki2010exact,come2013model}, where the different contexts allow to specify non-informative priors.
That being said, we provide here the clustering solutions for the three methods mentioned mainly to give a better understanding of the exact ICL method itself, and to show how sensitive to hyperparameters the 
solutions obtained can be, where model-based clustering optimal configurations are used as references.

\subsection{Dataset 1} 
\begin{figure}[htbp]
\centering
 \includegraphics[width=0.4\textwidth]{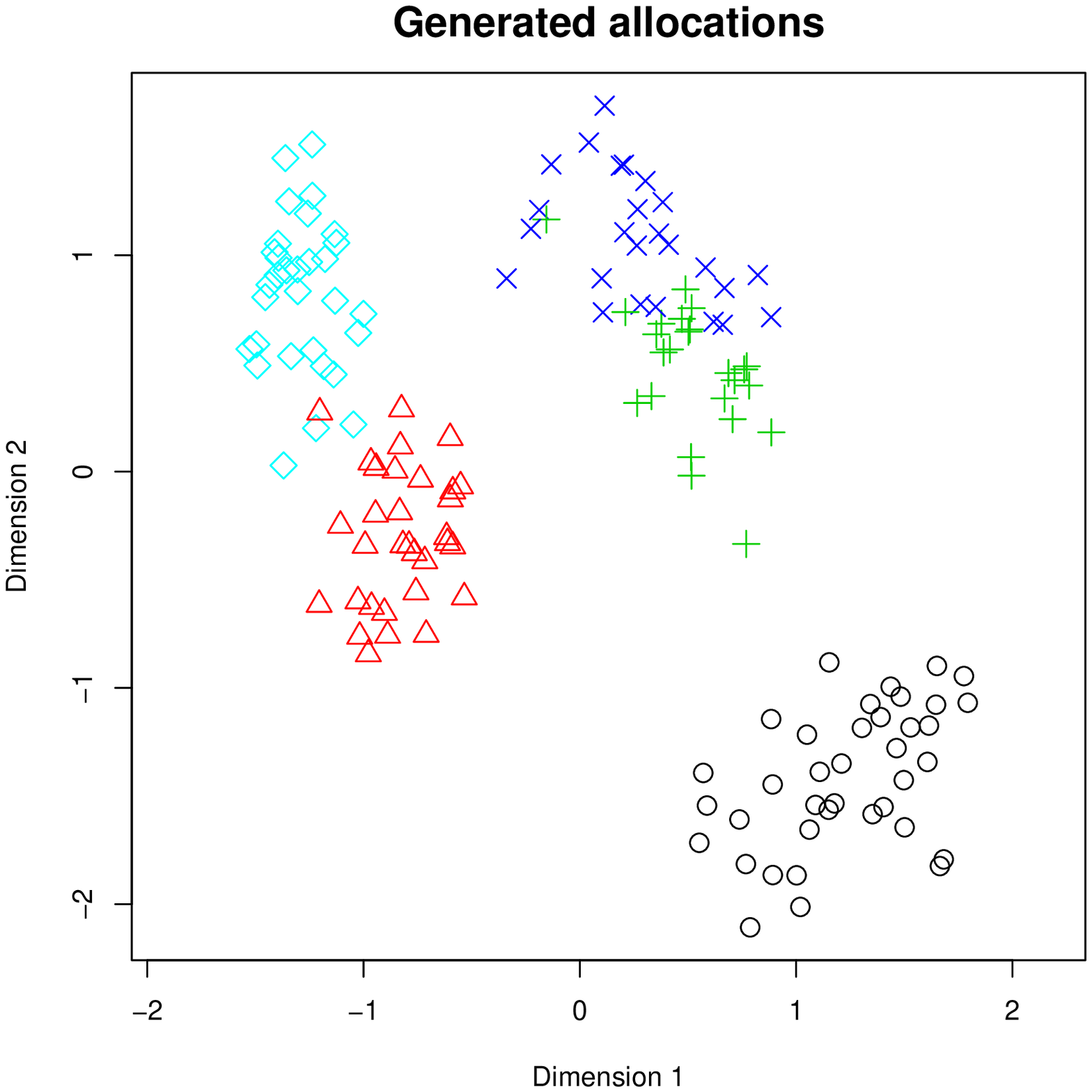}
 \includegraphics[width=0.4\textwidth]{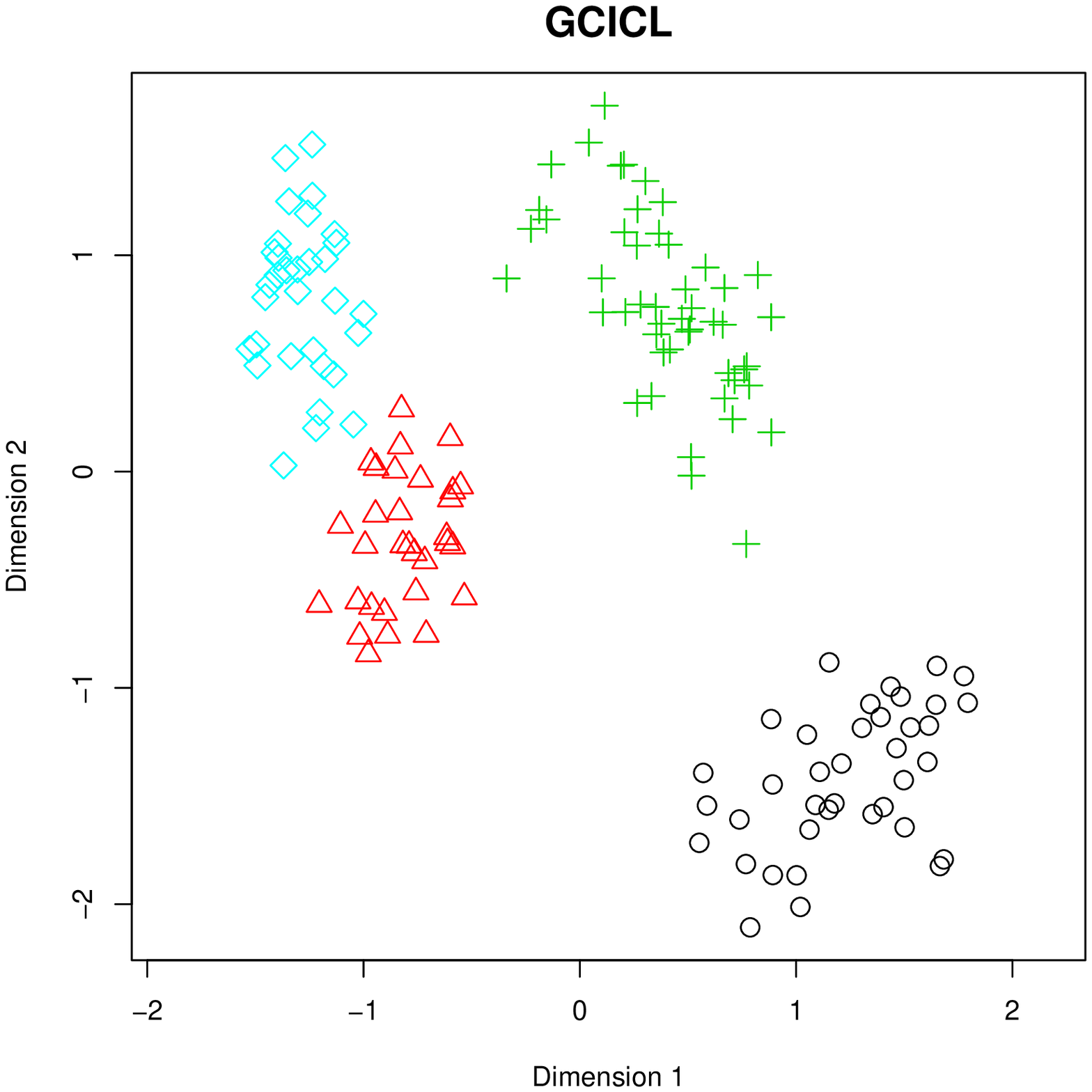}
 \includegraphics[width=0.4\textwidth]{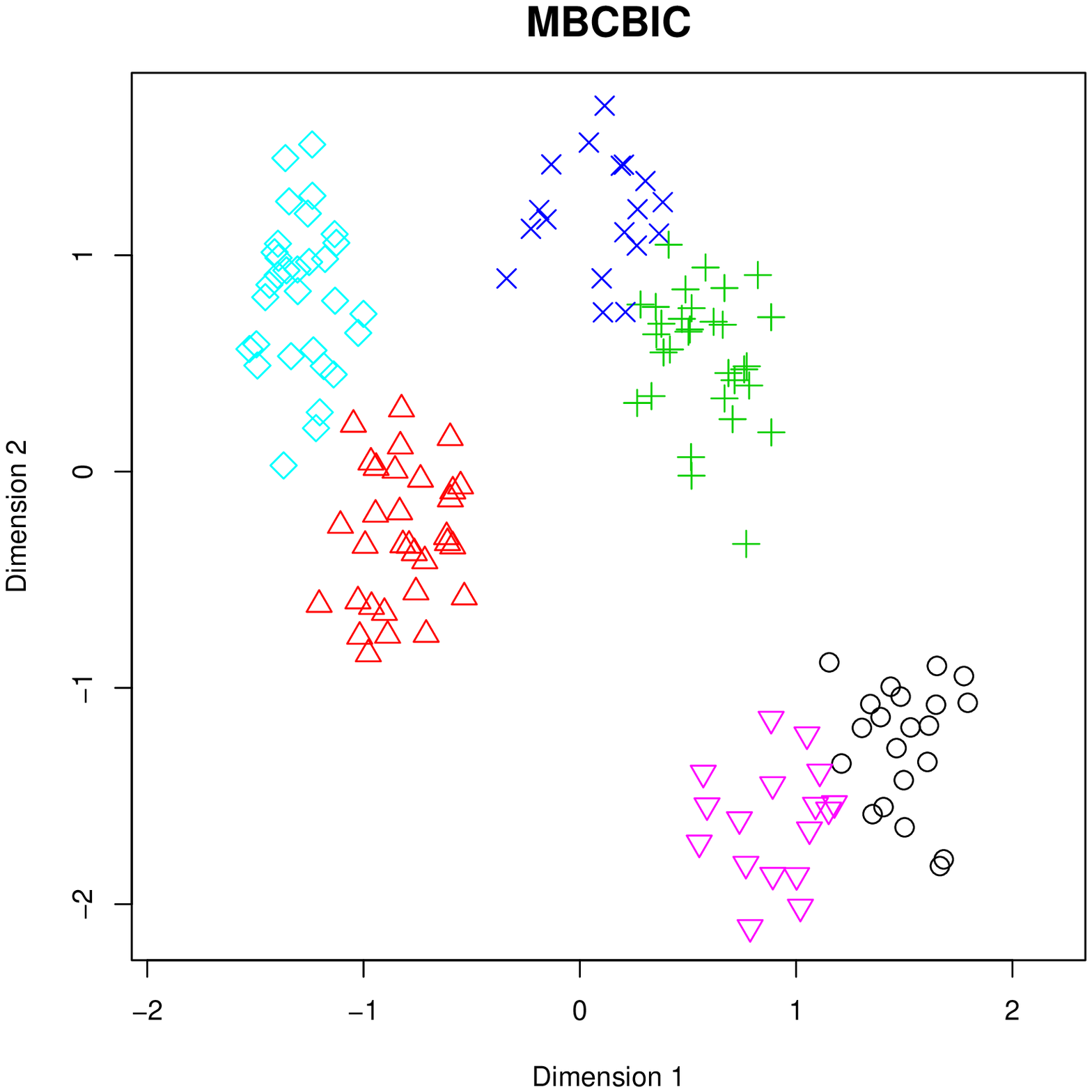}
 \includegraphics[width=0.4\textwidth]{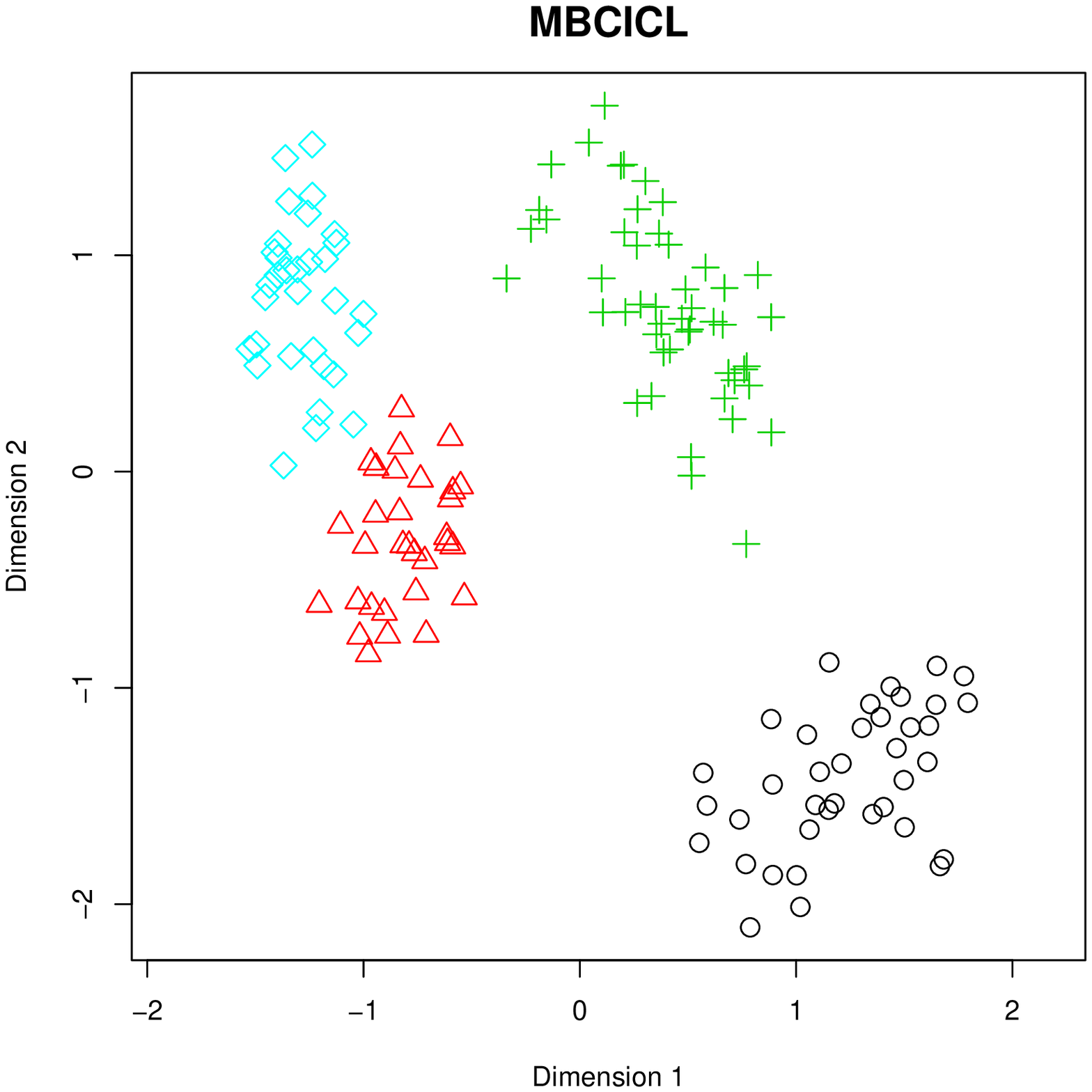}
 \caption{\textbf{Dataset 1.} On the top left, an image of the true allocations simulated from the likelihood model. On the top right image the configuration corresponding to the best $ICL_{ex}$ value is represented.
Images on the bottom row correspond instead to the standard model-based clustering routines (BIC on the left and ICL on the right).}
 \label{fig:Sim1}
\end{figure}

The first simulated dataset is mainly intended to describe the influence of the hyperparameters $\tau$ and $\omega$ on the corresponding GCICL solutions.
Here $150$ observations are realised from a $5$ components bivariate Gaussian mixture model. The hyperparameters used to generate the data are
$\alpha=10,\ \boldsymbol{\mu}=(0,0)^t,\ \nu=b+2,\ \tau=0.05$ and the identity matrix is chosen as $\boldsymbol{\xi}$. 
Figure \ref{fig:Sim1} represents the data with the generated allocations, and the optimal clustering solutions GCICL, MBCBIC and MBCICL.
It is important to stress that 
we do not really expect the GCICL to agree with the configuration simulated from the likelihood model. 
In this particular datasets, the MBCBIC and MBCICL both fail at capturing the ``true'' allocations generated.
The GCICL configuration represented in Figure \ref{fig:Sim1} corresponds to the best value of the $ICL_{ex}$ obtained among several possible choices of hyperparameters. 
The $ICL_{ex}$ values corresponding to other GCICL solutions are shown in Table \ref{table:Sim1}. 
It is clear that for this dataset the hyperparameters $\tau$ and $\omega$ both have a strong influence on the configuration returned.
\begin{table}[htb]
\centering
\begin{tabular}{cccc}
   \hline
& & $k$ & \\ 
   \hline
  \multicolumn{2}{c}{MBCBIC} & 6 & \\
  \multicolumn{2}{c}{MBCICL} & 4 & \\
   &  & \\
   \hline
$\tau$ & $\omega$ & $k$ & $ICL_{ex}$ \\ 
  \hline
0.1 & 0.1 &   5 & -304.68 \\ 
  0.1 &   1 &   3 & -313.28 \\ 
  0.1 &  10 &   2 & -416.91 \\ 
  0.01 & 0.1 &   4 & -299.29 \\ 
  0.01 &   1 &   3 & -317.08 \\ 
  0.01 &  10 &   2 & -420.95 \\ 
   \hline
\end{tabular}
\caption{\textbf{Dataset 1.} Output of the Greedy Combined ICL for different sets of hyperparameters. For all cases $\alpha=4,\ \nu=3$ and $\boldsymbol{\mu}=\left( 0,0 \right)^t$.
 The algorithm has been run $10$ times and $15$ iterations for each set of hyperparameters, with default parameters for $\beta_1$ and $\beta_2$.
 It can be noted that the hyperparameters $\tau$ and $\omega$, regulating the positioning of the points and cluster centres, can be very influential on the quality of the final solution.}
\label{table:Sim1}
\end{table}
It is also relevant that the best GCICL configuration proposed is here equivalent to the MBCICL solution, while the MBCBIC configuration estimates a larger number of groups.

\subsection{Dataset 2}
\begin{figure}[htb]
\centering
 \includegraphics[width=0.325\textwidth]{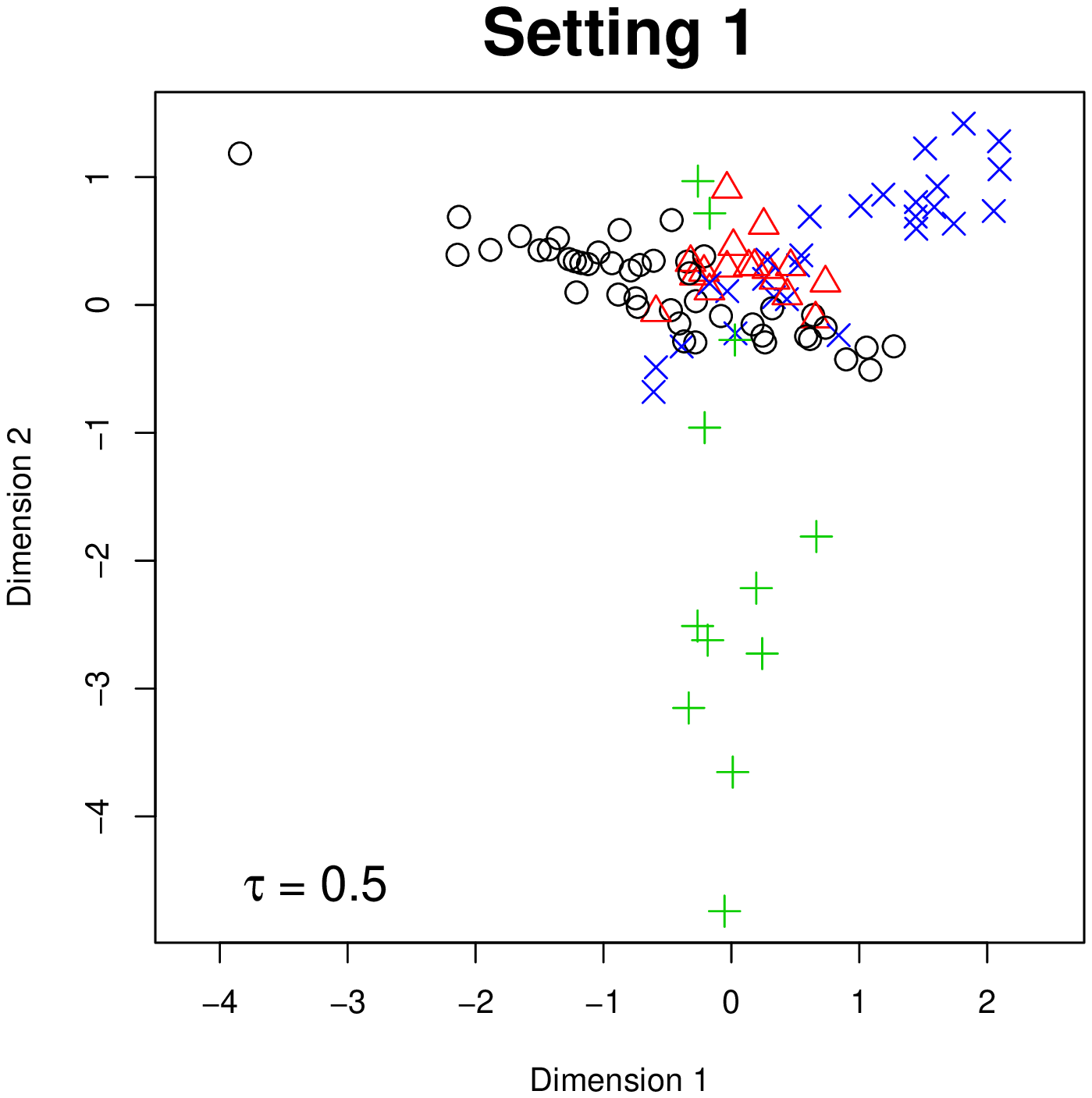}
 \includegraphics[width=0.325\textwidth]{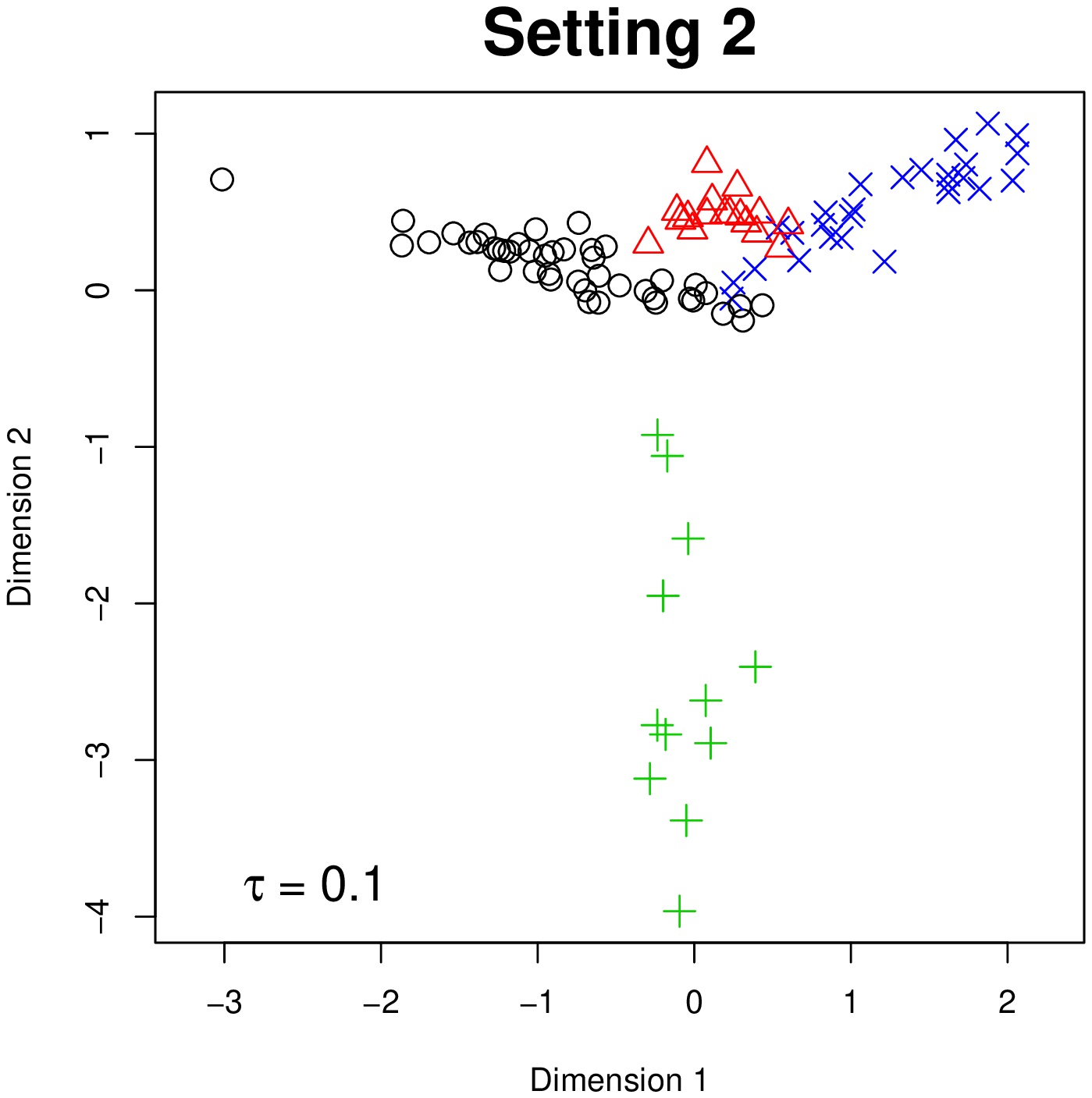}
 \includegraphics[width=0.325\textwidth]{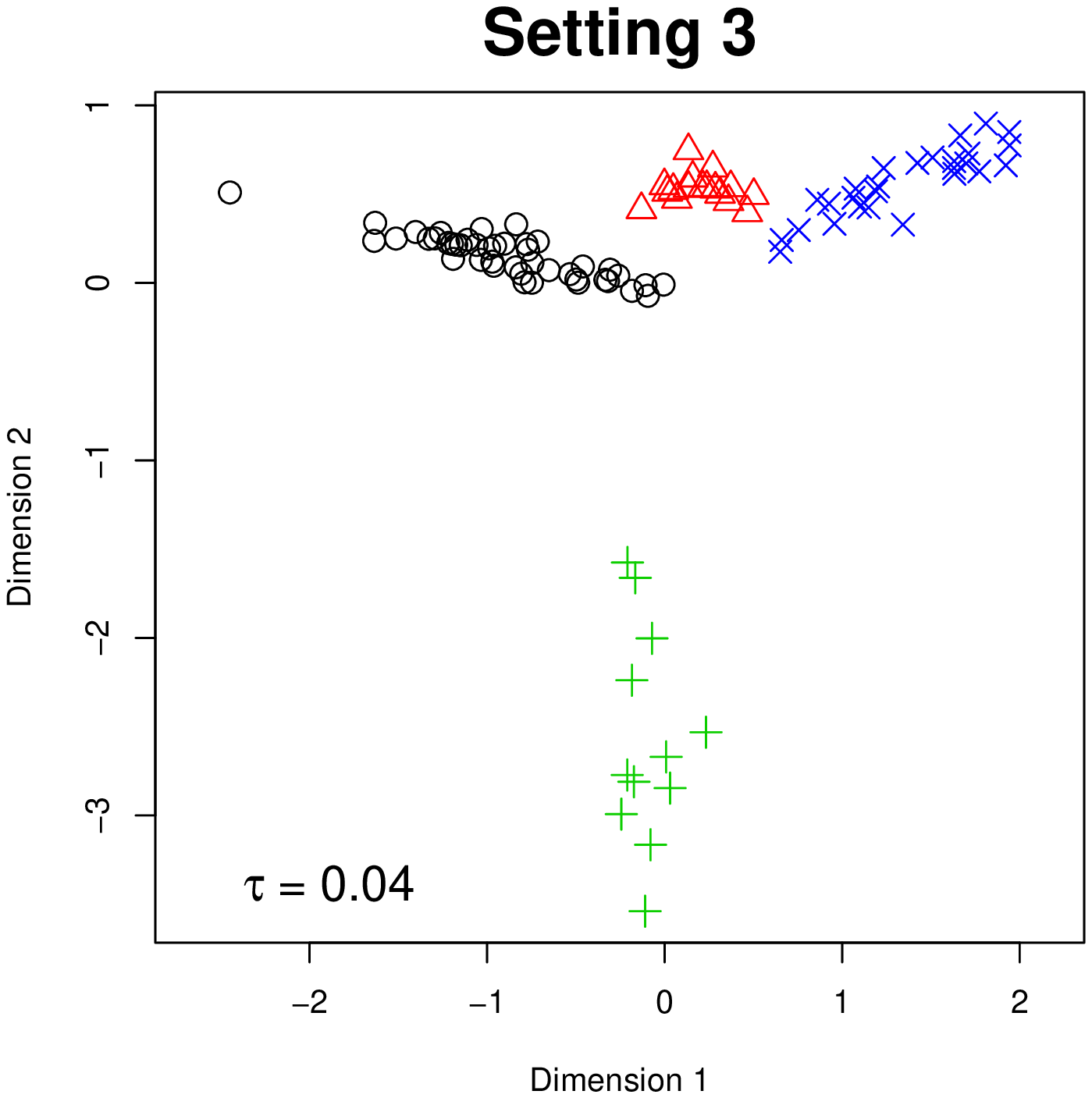}
 \includegraphics[width=0.325\textwidth]{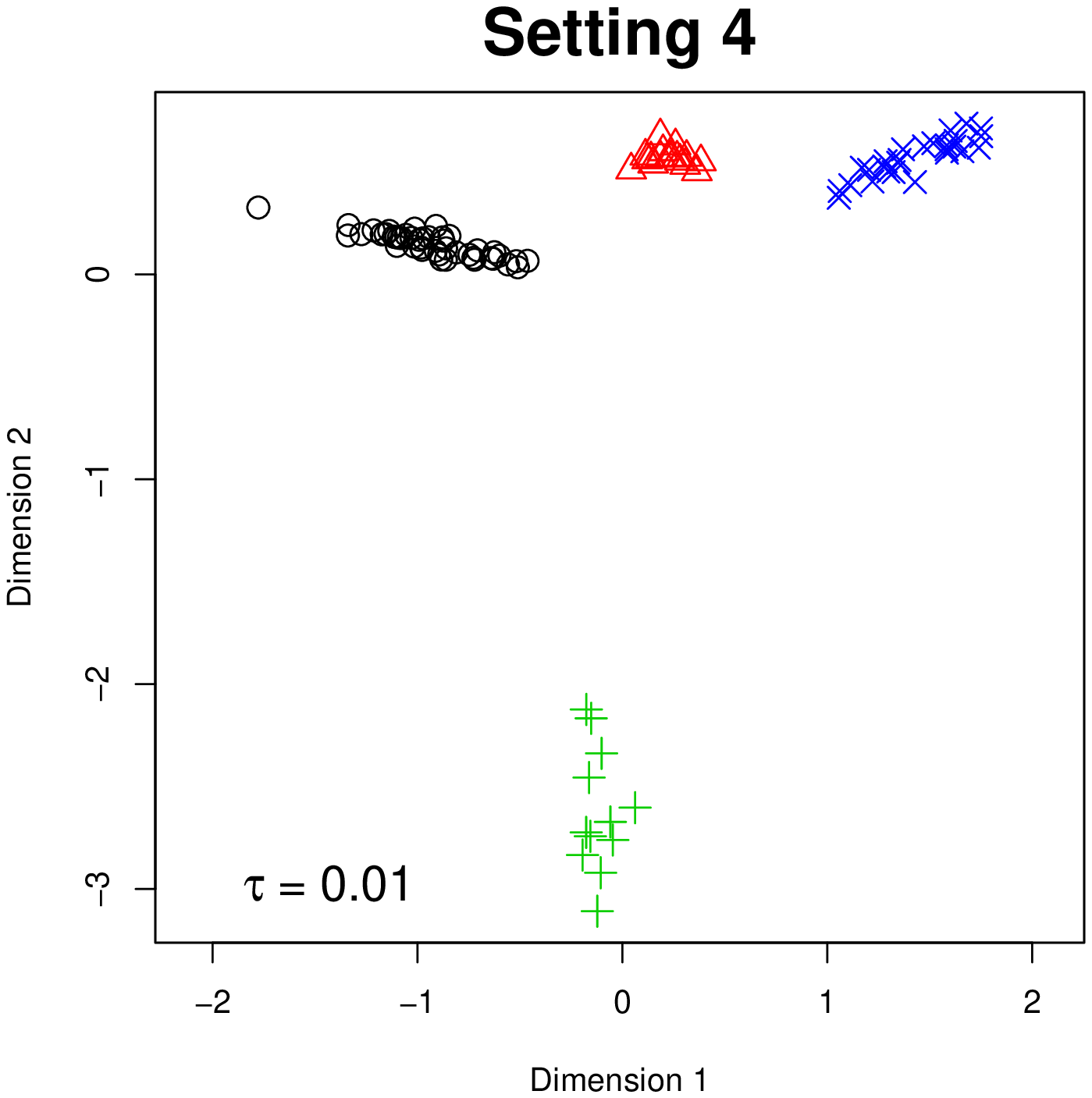}
 \includegraphics[width=0.325\textwidth]{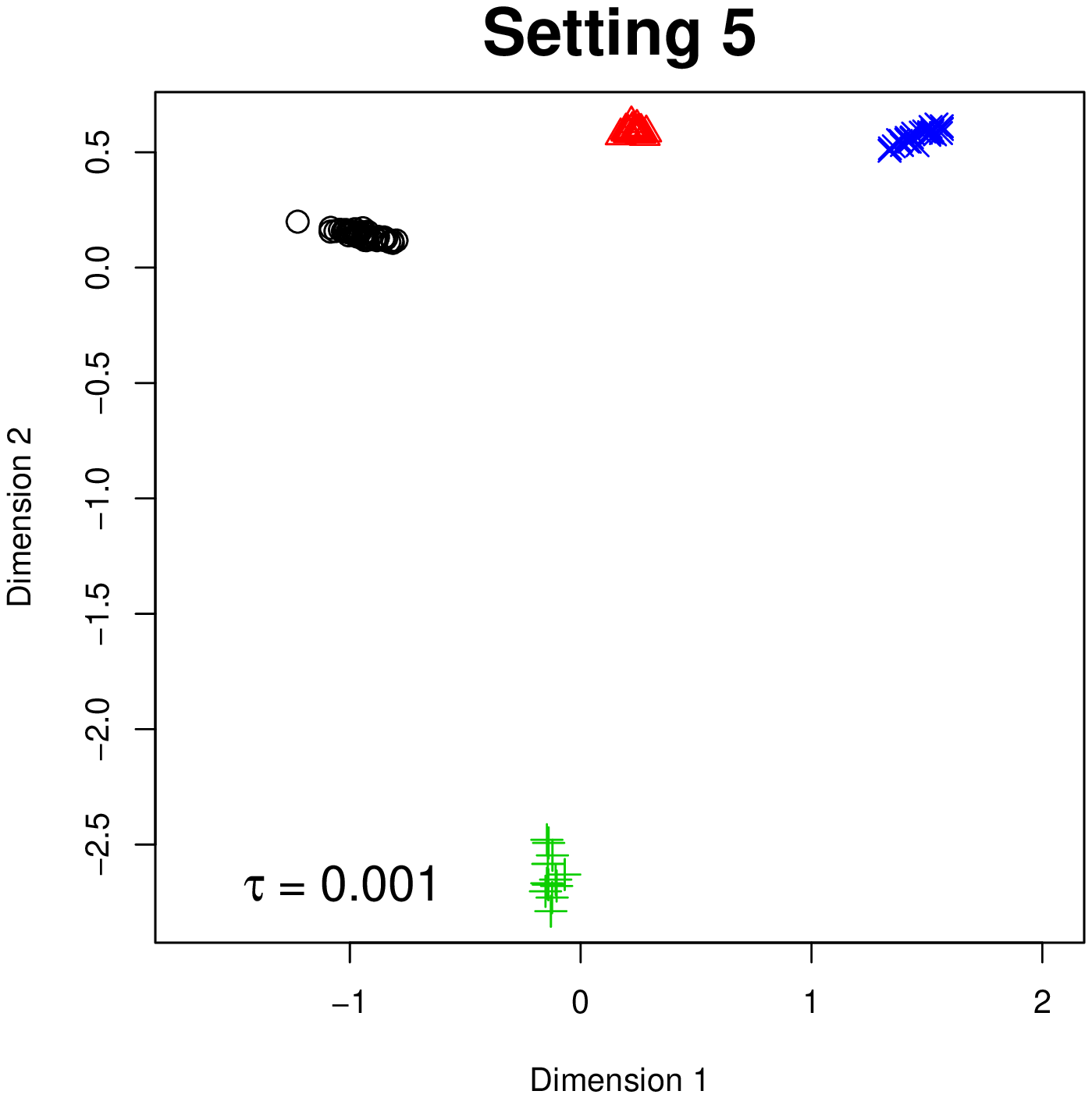}
 \caption{\textbf{Dataset 2.} The $5$ Settings used. True hyperparameters used for the generation are $\alpha=4,\ \boldsymbol{\mu}=(0,0)^t,\ \nu=3$ and $\omega=0.5$. With regard to $\tau$, the values $\left\{0.5,0.1,0.04,0.01,0.001 \right\}$ 
 have been used, each value corresponding to an image starting from top left to bottom right, by rows.}
 \label{fig:Sensitivity}
\end{figure}

\begin{figure}[htb]
\centering
 \includegraphics[width=0.325\textwidth]{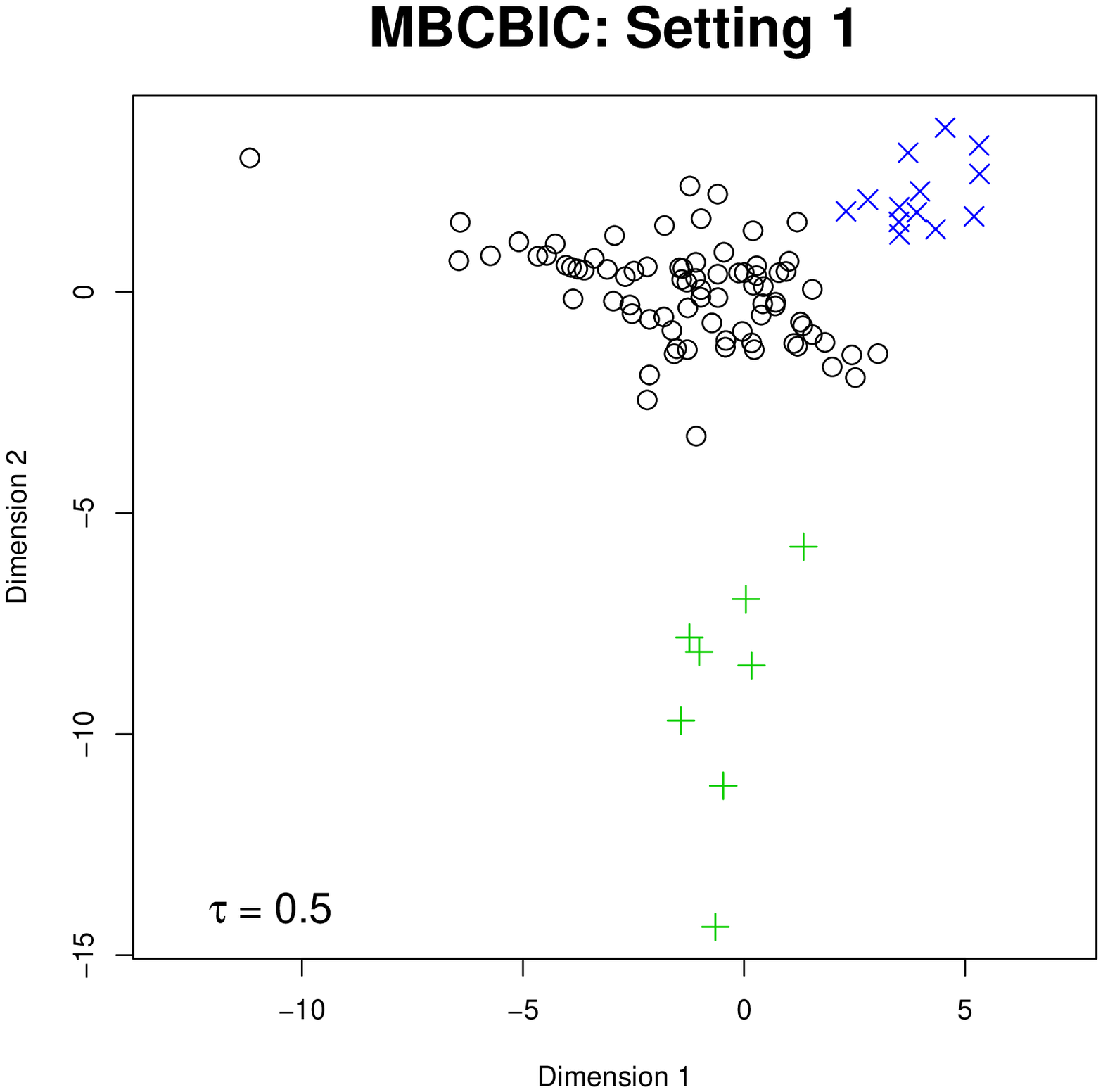}
 \includegraphics[width=0.325\textwidth]{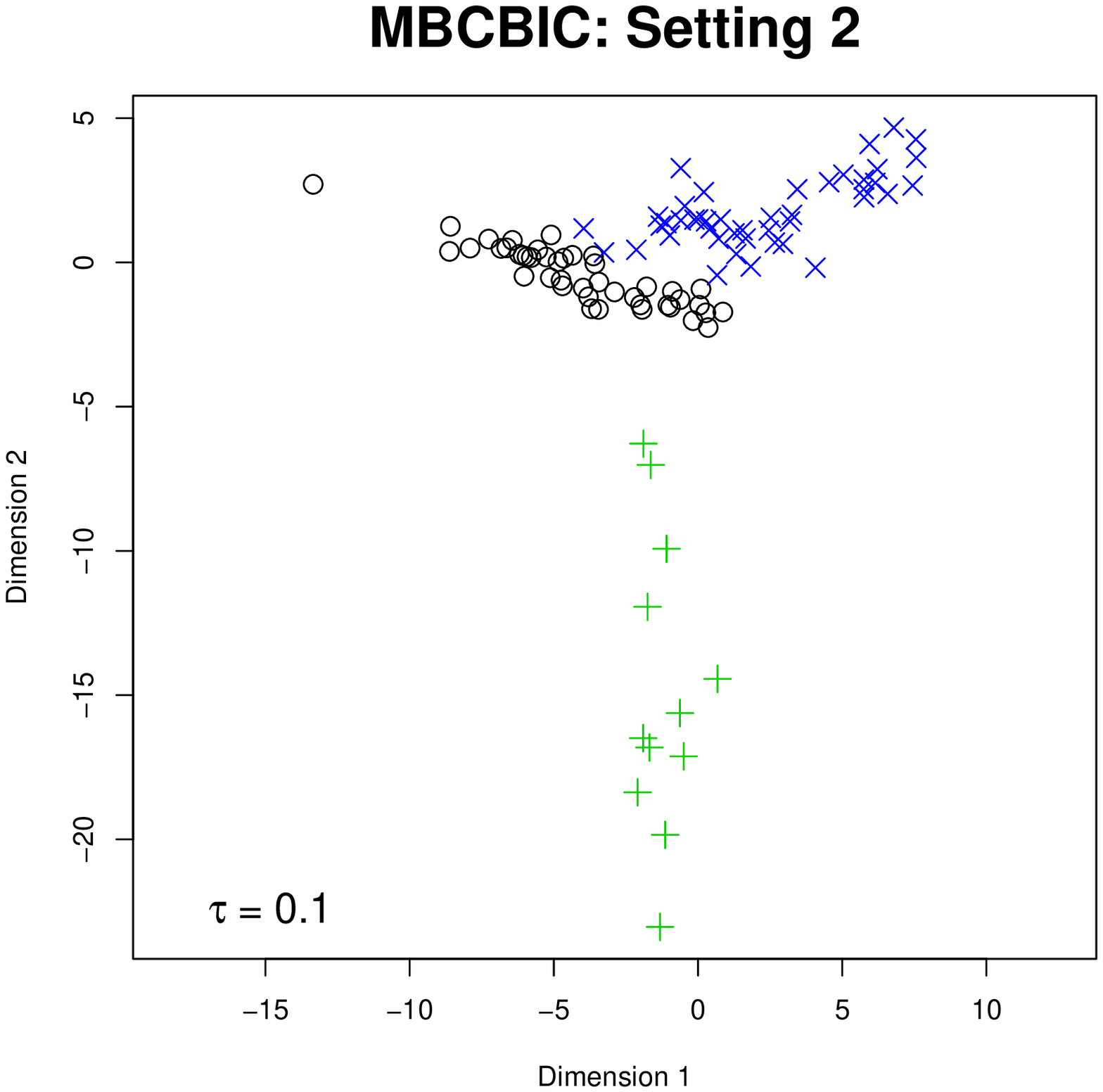}
 \includegraphics[width=0.325\textwidth]{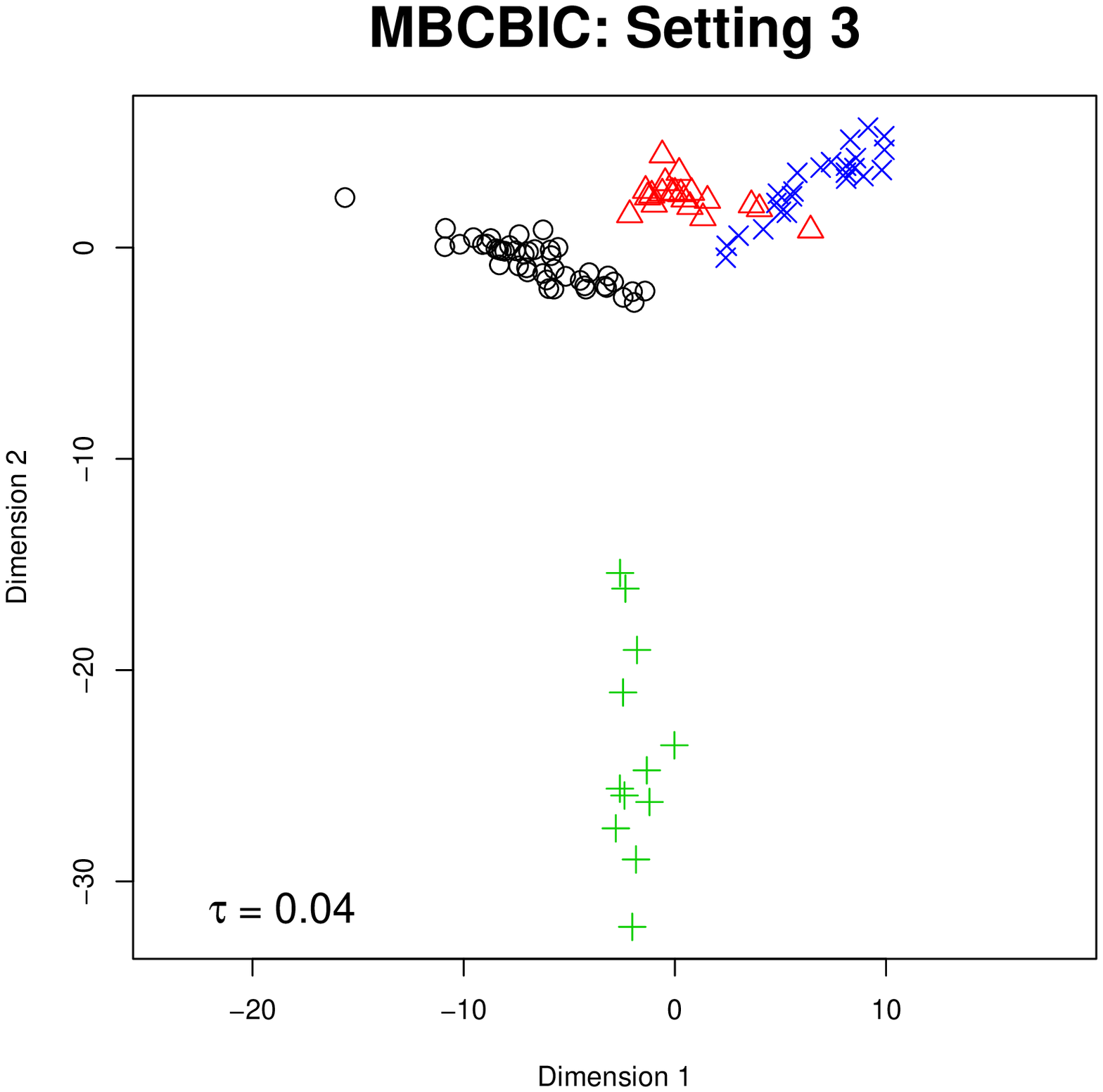}
 \includegraphics[width=0.325\textwidth]{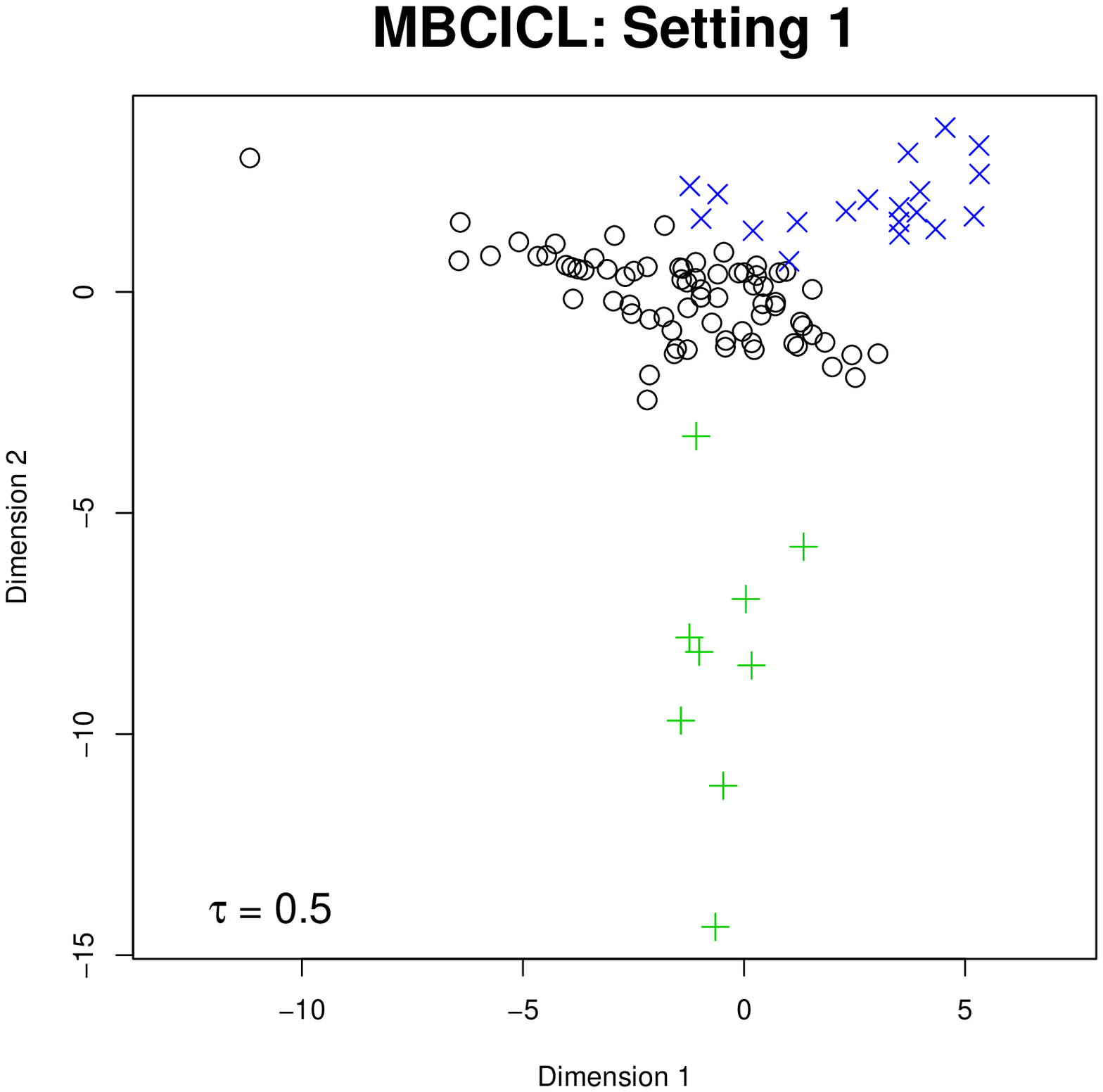}
 \includegraphics[width=0.325\textwidth]{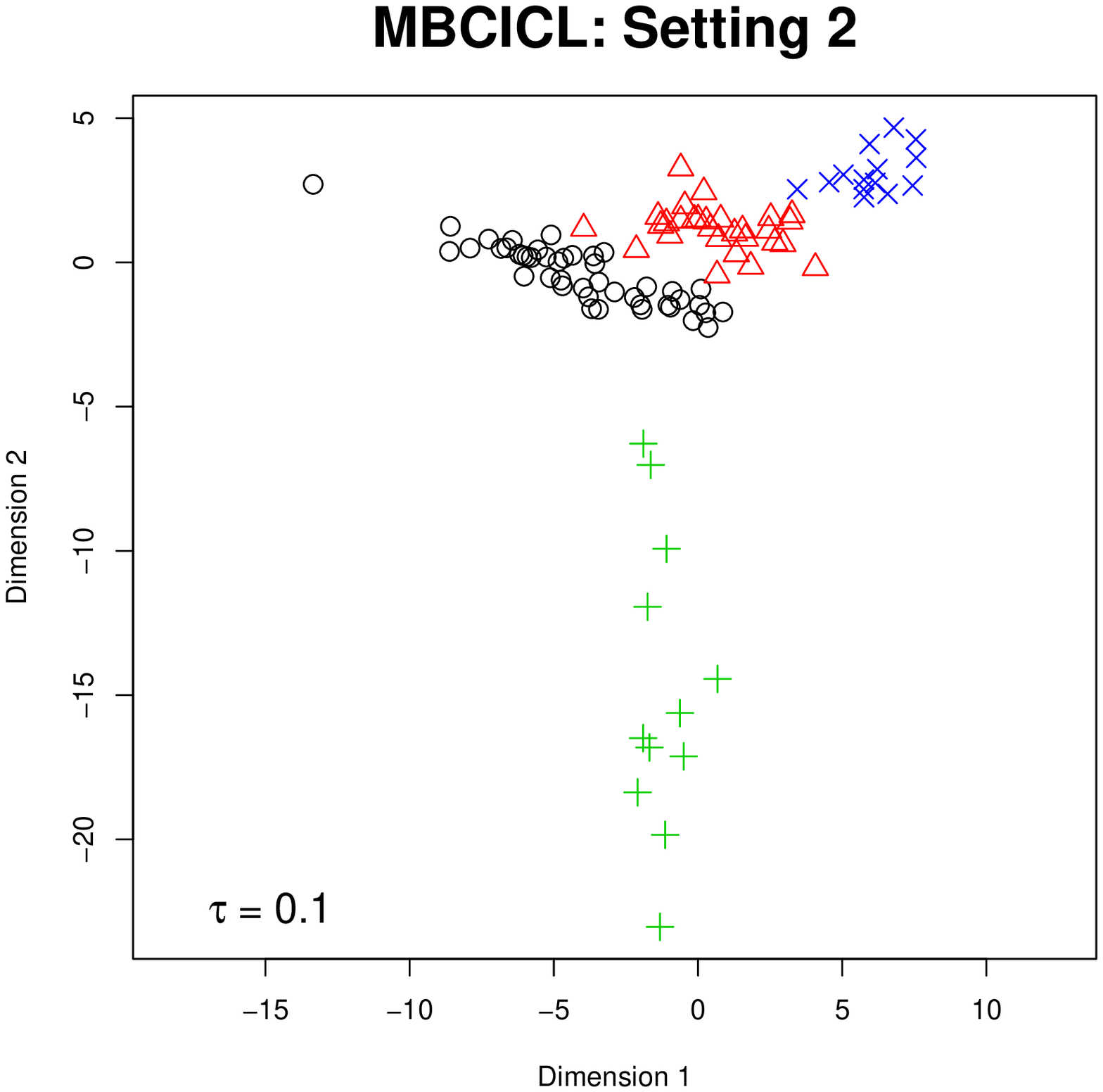}
 \includegraphics[width=0.325\textwidth]{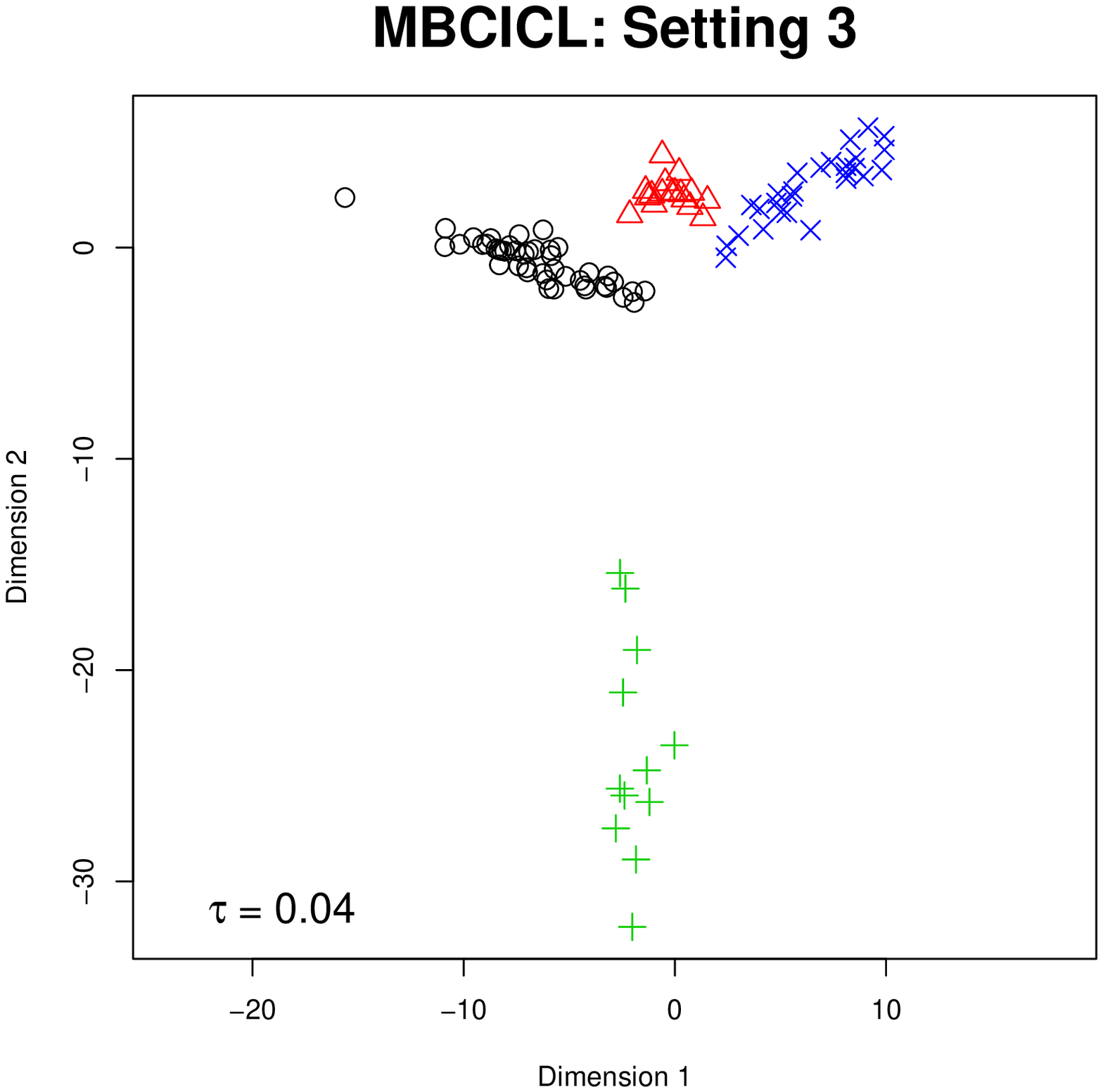}
 \caption{\textbf{Dataset 2.} Top row: MBCBIC optimal configurations corresponding to cases $\tau=0.5$ (left), $\tau=0.1$ (centre) and $\tau=0.04$ (right).
 Bottom row: MBCICL optimal configurations corresponding to cases $\tau=0.5$ (left), $\tau=0.1$ (centre) and $\tau=0.04$ (right). 
 ICL performs appreciably well even for the Setting 2 ($\tau=0.1$) while BIC recovers the generated allocations only for 
 Settings 4 and 5.}
 \label{fig:SensitivityMBC}
\end{figure}

Here we focus more on the role of the hyperparameters in determining the final GCICL solution, essentially providing a sensitivity analysis.
In particular we make use of simulated data to show how the performance of the Greedy Combined ICL in choosing the number of groups changes when different degrees of separation between clusters are exhibited.
We use $5$ datasets simulated using the bivariate Gaussian mixture model described in Section \ref{sec:MixtureModels}. The hyperparameters used to generate the different datasets are exactly the same in each case, 
with the exception of $\tau$, which takes $5$ different values. Since $\tau$ describes the distance between cluster centres, we essentially obtain $5$ equivalent datasets, that differ one from the other only with respect to the 
distance between the ``real'' generated cluster centres. In this way, we can highlight the performances of our algorithm under different levels of group overlapping. The datasets with the allocations simulated from the likelihood models are 
represented in Figure \ref{fig:Sensitivity}. The most relevant cases for the MBC routines are represented in Figure \ref{fig:SensitivityMBC}.

The Greedy Combined ICL is run for several choices of hyperparameters, providing a sensitivity analysis with respect to the estimated number of groups.
The complete results are provided in Table \ref{table:sensitivity}.
\begin{table}[htb]
\footnotesize
\centering
\begin{tabular}{|c|c|c|c|c|c|}
   \cline{2-6}
 \multicolumn{1}{c}{} & \multicolumn{5}{|c|}{\textbf{Number of groups}} \\ 
   \cline{2-6}
 \multicolumn{1}{c|}{}& Setting 1& Setting 2& Setting 3& Setting 4& Setting 5 \\ 
   \hline
  MBCBIC &   4 &    3 &    4  &   4 &    4  \\ 
   \hline
  MBCICL &   4 &    4 &    4  &   4 &    4  \\ 
   \hline
  \multicolumn{5}{c}{}\\
\end{tabular}\\
\begin{tabular}{|ccc|cc|cc|cc|cc|cc|}
  \hline
 \multicolumn{13}{|c|}{\textbf{GCICL}} \\ 
  \hline
\multicolumn{3}{|c|}{Hyperparameters} & \multicolumn{2}{c|}{Setting 1} & \multicolumn{2}{c|}{Setting 2} & \multicolumn{2}{c|}{Setting 3} & \multicolumn{2}{c|}{Setting 4} & \multicolumn{2}{c|}{Setting 5} \\ 
  \hline
$\omega$ & $\tau$ & $\alpha$ & $k$ & $ICL_{ex}$& $k$ & $ICL_{ex}$& $k$ & $ICL_{ex}$& $k$ & $ICL_{ex}$& $k$ & $ICL_{ex}$ \\ 
  \hline
0.1 & 0.1 & 0.5 &   3 & -470.75 &   3 & -520.48 &   4 & -532.61 &   4 & -545.19 &   4 & -610.77 \\ 
  0.1 & 0.1 &   4 &   5 & -481.55 &   4 & -519.79 &   4 & -530.63 &   4 & -543.22 &   4 & -608.80 \\ 
  0.1 & 0.1 &  10 &   6 & -487.85 &   \textcolor{red}{5} & \textcolor{red}{-530.35} &   4 & -531.26 &   4 & -543.85 &   4 & -609.43 \\ 
  0.1 & 0.01 & 0.5 &   4 & -483.81 &   4 & -526.73 &   4 & -537.43 &   4 & -539.06 &   4 & -554.99 \\ 
  0.1 & 0.01 &   4 &   3 & -478.08 &   4 & -525.57 &   4 & -535.46 &   4 & -537.09 &   4 & -553.02 \\ 
  0.1 & 0.01 &  10 &   5 & -489.40 &   \textcolor{red}{5} & \textcolor{red}{-538.86} &   4 & -536.09 &   4 & -537.72 &   4 & -553.65 \\ 
    1 & 0.1 & 0.5 &   3 & -456.89 &   4 & -498.67 &   4 & -510.39 &   4 & -522.37 &   4 & -586.21 \\ 
    1 & 0.1 &   4 &   5 & -462.65 &   4 & -497.09 &   4 & -508.42 &   4 & -520.40 &   4 & -584.24 \\ 
    1 & 0.1 &  10 &   6 & -469.49 &   5 & -504.96 &   \textcolor{red}{5} & \textcolor{red}{-518.63} &   4 & -521.03 &   4 & -584.87 \\ 
    1 & 0.01 & 0.5 &   3 & -458.84 &   4 & -505.01 &   4 & -515.45 &   4 & -516.99 &   4 & -532.13 \\ 
    1 & 0.01 &   4 &   5 & -471.48 &   4 & -503.43 &   4 & -513.48 &   4 & -515.02 &   4 & -530.16 \\ 
    1 & 0.01 &  10 &   5 & -473.13 &   5 & -511.30 &   4 & -514.11 &   \textcolor{red}{5} & \textcolor{red}{-526.20} &   4 & -530.79 \\ 
   10 & 0.1 & 0.5 &   3 & -455.50 &   3 & -508.42 &   4 & -521.38 &   4 & -530.16 &   4 & -582.66 \\ 
   10 & 0.1 &   4 &   3 & -457.49 &   4 & -507.28 &   4 & -519.41 &   4 & -528.20 &   4 & -580.69 \\ 
   10 & 0.1 &  10 &   3 & -462.98 &   4 & -508.85 &   4 & -520.04 &   4 & -528.83 &   4 & -581.32 \\ 
   10 & 0.01 & 0.5 &   3 & -461.30 &   3 & -515.03 &   4 & -527.59 &   4 & -528.68 &   4 & -539.72 \\ 
   10 & 0.01 &   4 &   3 & -463.33 &   3 & -513.43 &   4 & -525.62 &   4 & -526.72 &   4 & -537.75 \\ 
   10 & 0.01 &  10 &   4 & -481.63 &   4 & -515.51 &   4 & -526.25 &   4 & -527.35 &   4 & -538.38 \\
   \hline
\end{tabular}
\caption{\textbf{Dataset 2.} Sensitivity analysis for the simulated data. For each of the $5$ cases, the number of groups and the final best $ICL_{ex}$ values are reported. Hyperparameters $\nu$ and $\boldsymbol{\mu}$ are
set to default values. The number of iterations is $15$ while the number of repetitions is $10$.}
\label{table:sensitivity}
\end{table}
Note that even though the number of groups is always recovered exactly in the MBCICL, the corresponding clustering solution is not equivalent to the true one, indeed the groups have different shapes and sizes.
Essentially, both the MBCBIC and MBCICL recover the true structure of the data in the cases $3$, $4$ and $5$.
In the fifth setting the groups are so separated that all of the methods return exactly the generated classification. As concerns the intermediate situations, 
different outcomes are obtained. 
In all but $4$ hyperparameters' choices (coloured in red in the table), the Greedy Combined ICL algorithm returned a value for the $ICL_{ex}$ higher than the one provided by the generated allocations (evaluated under the same hyperparameters),
suggesting an appreciable rate of convergence.
Of course, using different initialisation and re-running the routine may help recover the global optimum for $ICL_{ex}$. 
But we want to use this example to emphasise that our Greedy Combined ICL approach is a heuristic algorithm, and thus it is not guaranteed to reach an optimal solution under the design chosen.
The Greedy ICL has also been run on the same framework (results are not shown here) obtaining no better results than the Greedy Combined ICL.
When the GCICL solutions are compared to the MBCBIC and MBCICL ones, we can essentially state that some choices of hyperparameters yield a rather different
optimal clustering, in particular when groups are overlapping (first two cases).
It seems indeed that $\alpha$, $\tau$ and $\omega$ can all affect the final solution considerably when groups are not well separated.
For each setting some choices of hyperparameters yield a meaningful solution that is reasonably close to the one we were expecting, while some sets return apparently inexplicable configurations.
This emphasises that, in some clustering problems, hyperparameters choice can be overwhelming.

\subsection{Dataset 3}
\begin{figure}[htb]
\centering
 \includegraphics[width=0.325\textwidth]{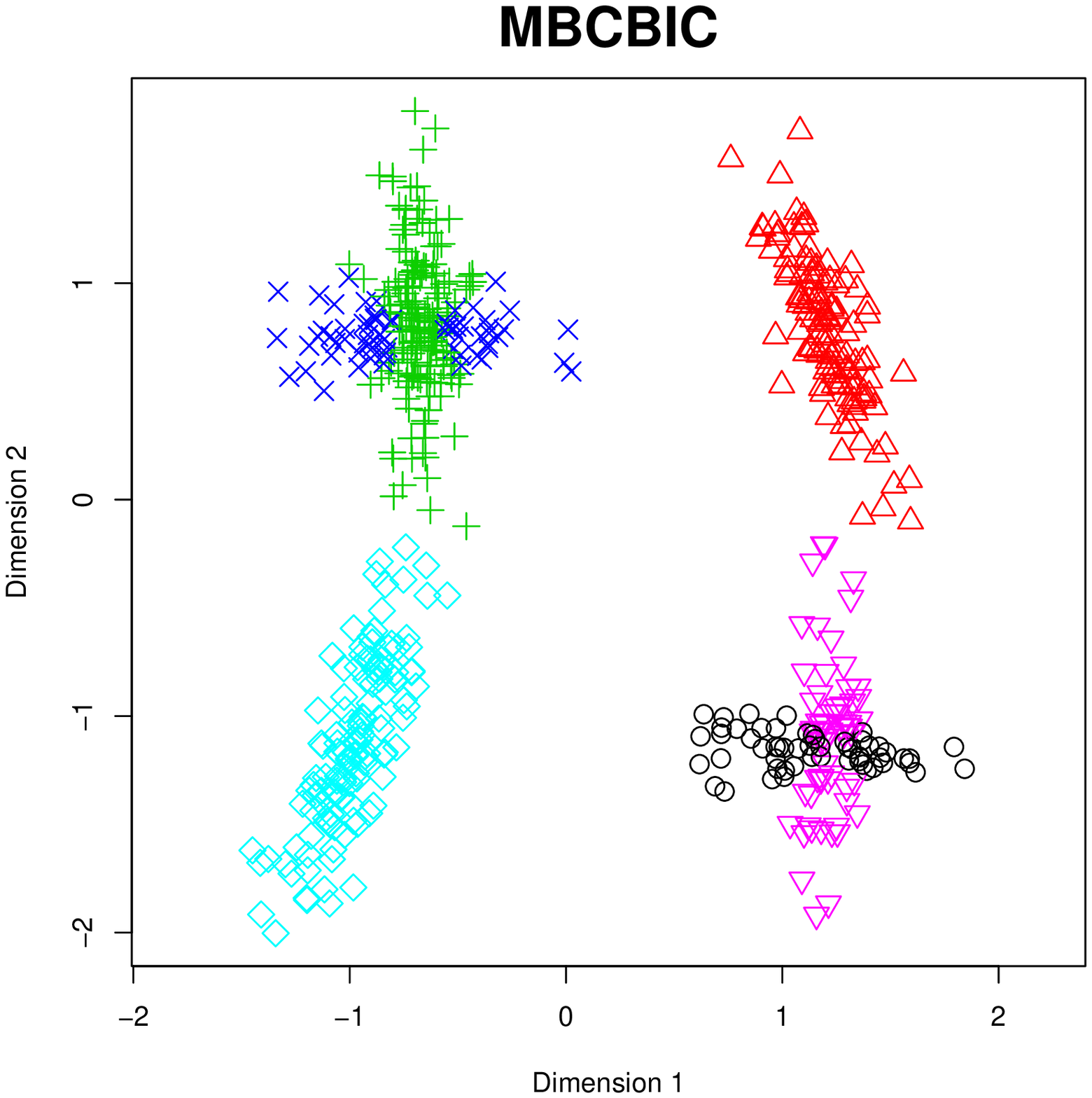}
 \includegraphics[width=0.325\textwidth]{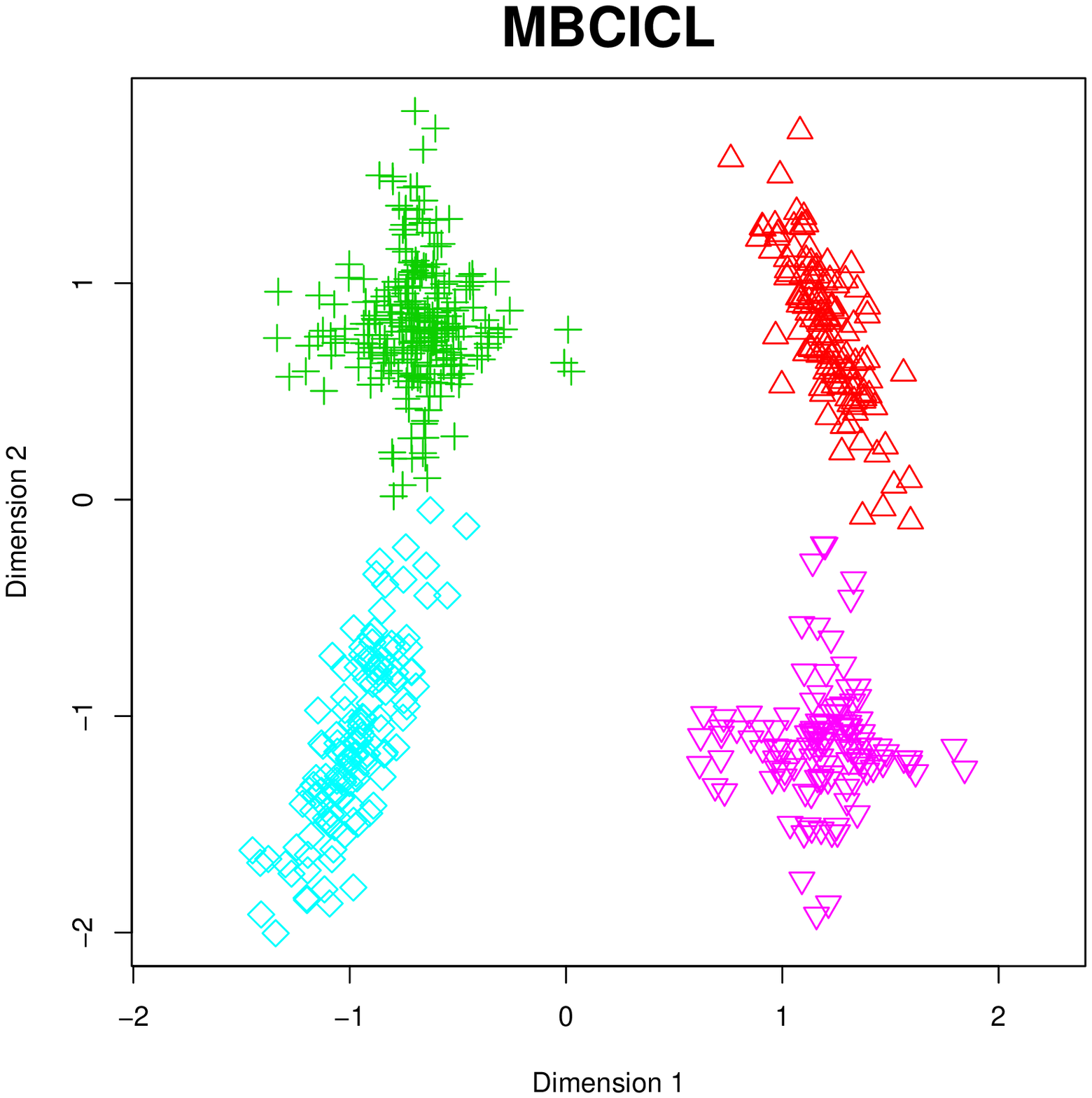}
 \includegraphics[width=0.325\textwidth]{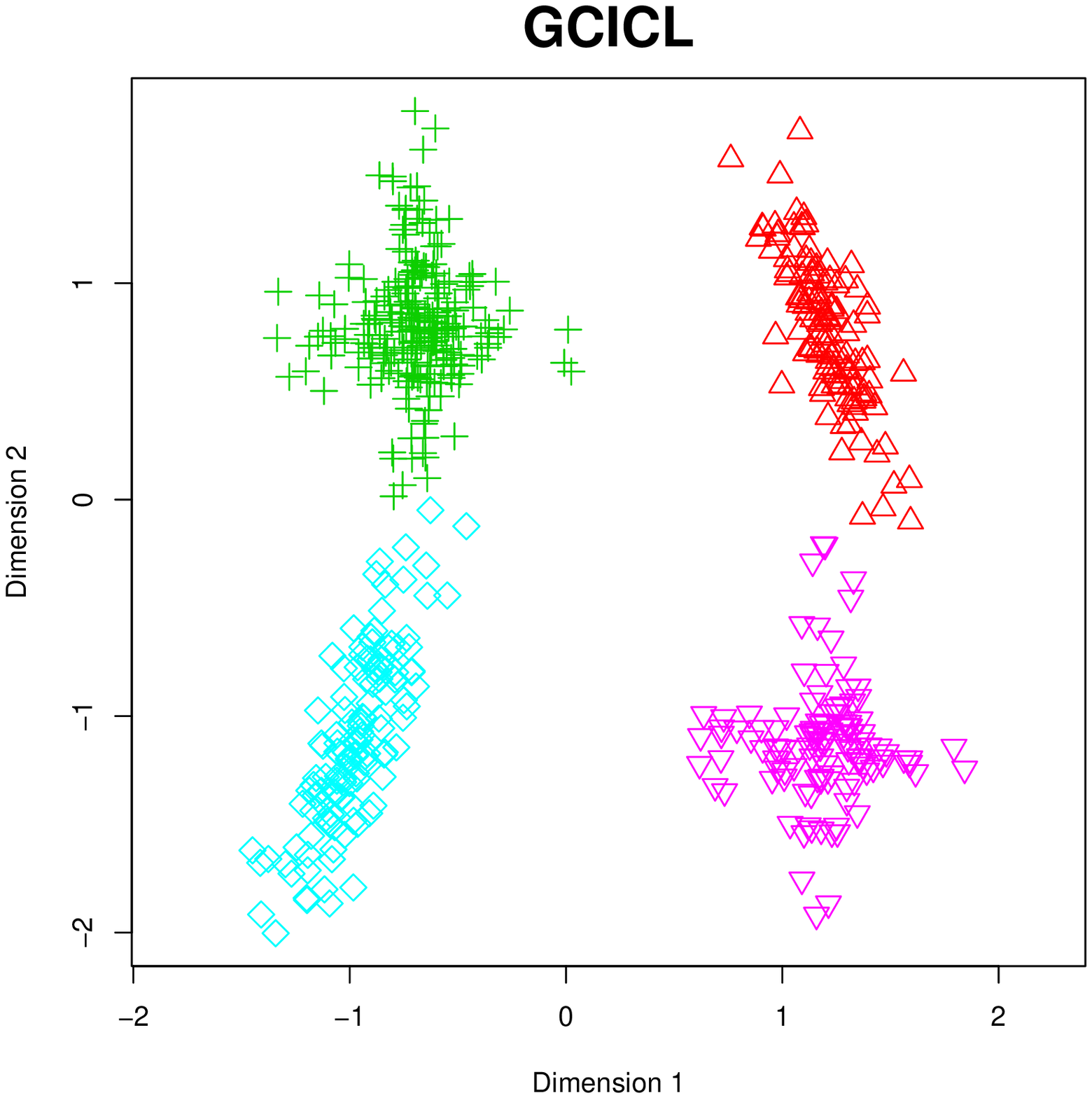}
 \caption{\textbf{Dataset 3.} Best configurations found: the MBCBIC solution (left), the MBCICL one (centre) and the GCICL configuration corresponding to the highest $ICL_{ex}$ value (right).}
 \label{fig:baudry}
\end{figure}

The data used in this experiment has been borrowed from the first simulated example of \textcite{baudry2010combining}.
Here $600$ points are realised from a $6$ components mixture of bivariate Gaussians, and groups exhibit strong overlapping:
observations are essentially divided in $4$ separated clusters, where $2$ clusters are made of $2$ components and the other $2$ clusters are made of a single component.
In \textcite{baudry2010combining} such dataset is used to highlight the different behaviour of BIC and ICL with respect to overlapping groups. While BIC tends to estimate the number of components exactly,
ICL penalises solutions exhibiting overlapping groups, thus underestimating the number of components, but unveiling a proper number of clusters.
We extend here the experiment by showing the results obtained through the Greedy Combined ICL on the same dataset. 
The algorithm is run for different sets of hyperparameters. The final number of groups and $ICL_{ex}$ values are reported in Table \ref{table:baudry}.
\begin{table}[htb]
\centering
\begin{tabular}{cccc}
   \hline
& & $k$ & \\ 
   \hline
  \multicolumn{2}{c}{MBCBIC} & 6 & \\
  \multicolumn{2}{c}{MBCICL} & 4 & \\
   &  & \\
  \hline
$\tau$ & $\omega$ & $k$ & $ICL_{ex}$ \\ 
  \hline
0.1 & 0.1 &   4 & -770.37 \\ 
  0.1 &   1 &   4 & -845.08 \\ 
  0.1 &  10 &   4 & -1281.61 \\ 
  0.01 & 0.1 &   4 & -764.57 \\ 
  0.01 &   1 &   4 & -845.69 \\ 
  0.01 &  10 &   4 & -1287.96 \\ 
  \hline
\end{tabular}
\caption{\textbf{Dataset 3.} Best number of groups and $ICL_{ex}$ values corresponding to different sets of hyperparameters. For each case, the Greedy Combined ICL algorithm has been run $10$ times for $10$ iterations, using as
parameters $\beta_1=0.2$ and $\beta_2=0.04$, in order to improve convergence for the larger dataset. Hyperparameters $\alpha$, $\nu$, and $\boldsymbol{\mu}$ are set to default values.
For this experiment, the optimal GCICL solutions are not really sensitive to different values of the hyperparameters.}
\label{table:baudry}
\end{table}
It is clear that different choices of hyperparameters do not affect the final estimation of the number of clusters, showing an agreement with the already known ICL solution of \textcite{baudry2010combining}.
The actual classification of the points for the different methods is shown in Figure \ref{fig:baudry}.
Apparently, in this dataset the sensitivity of the final clustering with respect to the hyperparameters is less strong, probably due to an increased number of observations and to the fact that the $4$ clusters are well separated.

\section{Galaxy dataset}\label{sec:RealDataApplications}
\begin{figure}[htb]
\centering
 \includegraphics[width=0.325\textwidth]{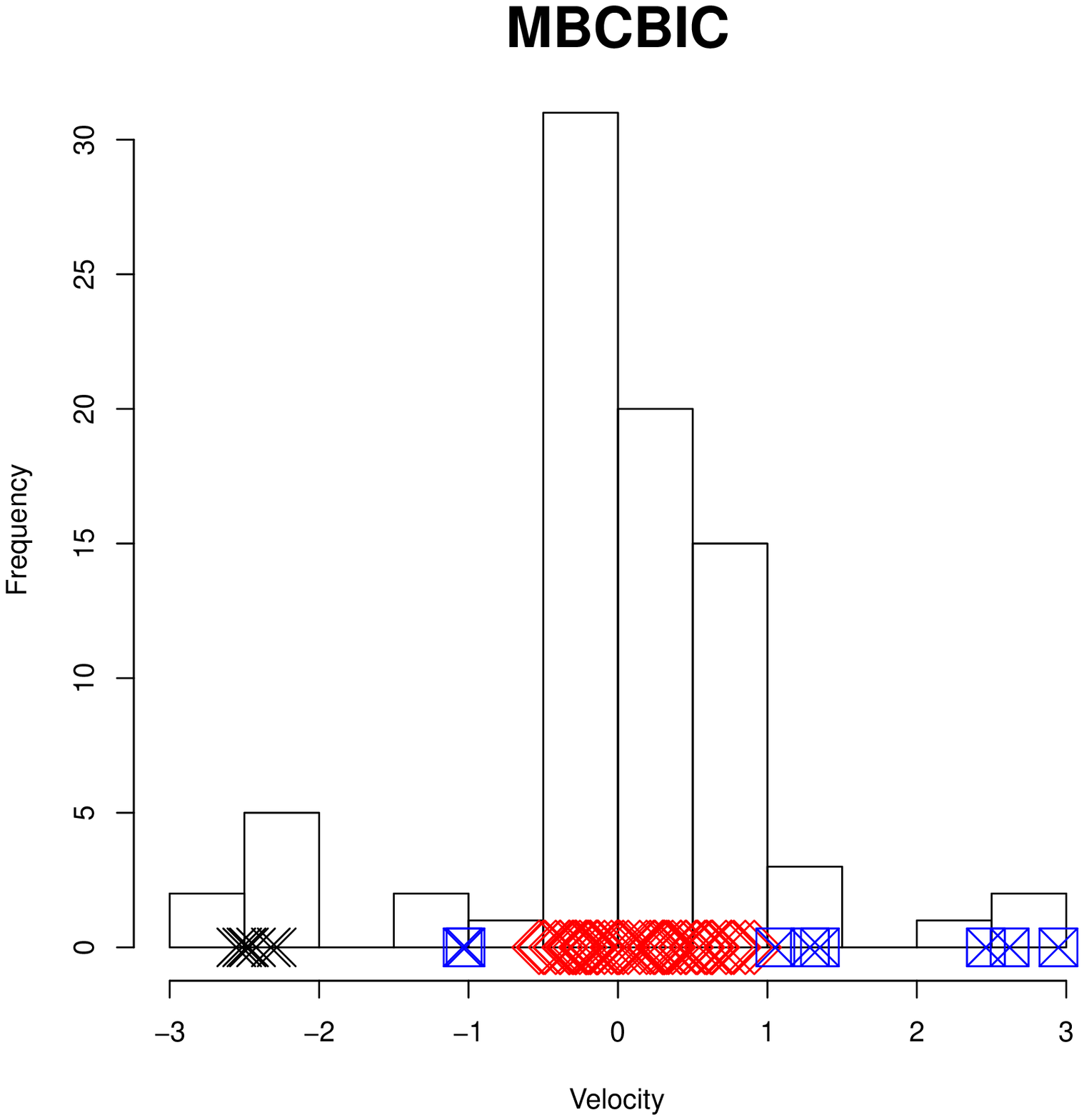}
 \includegraphics[width=0.325\textwidth]{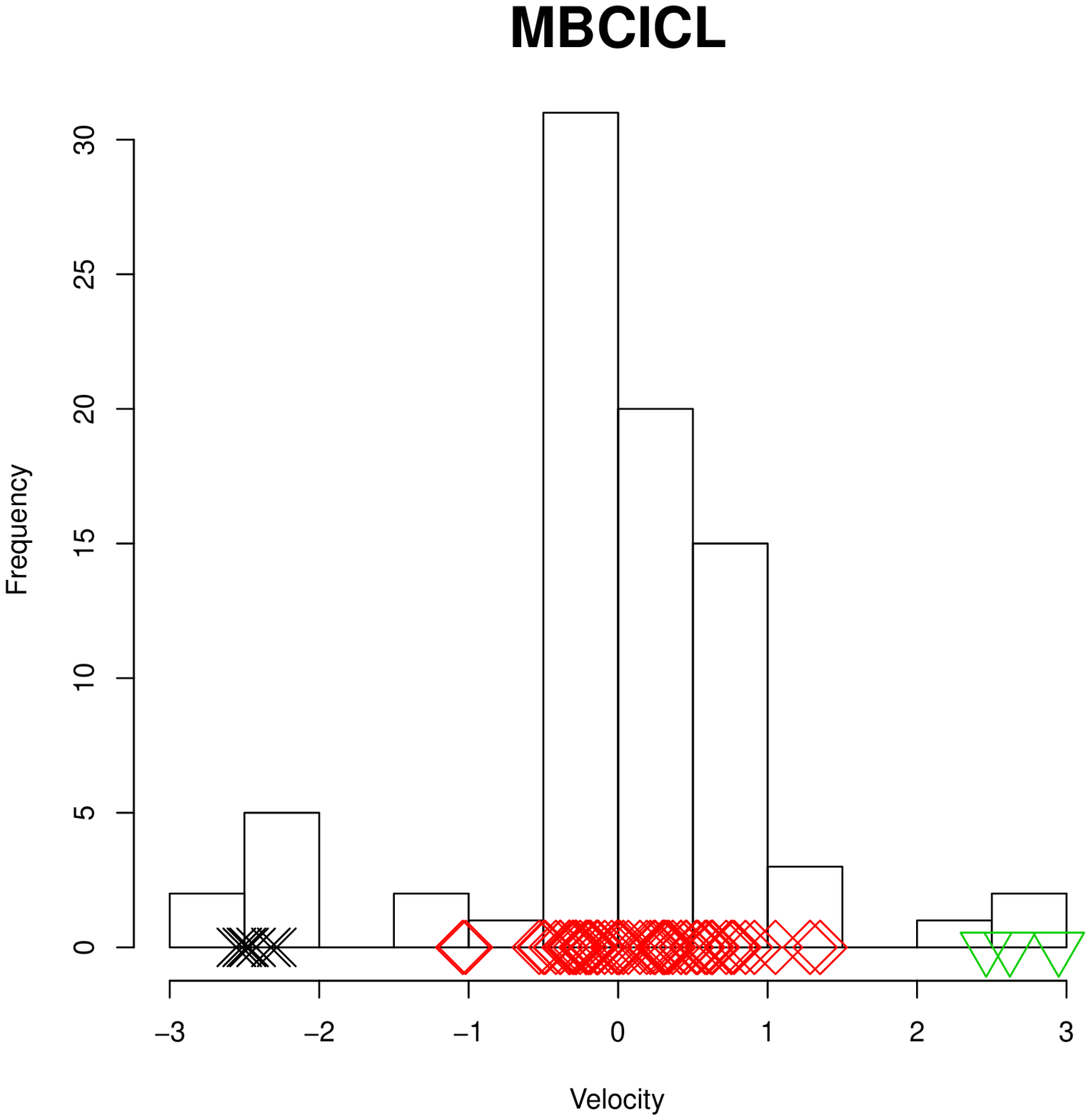}
 \includegraphics[width=0.325\textwidth]{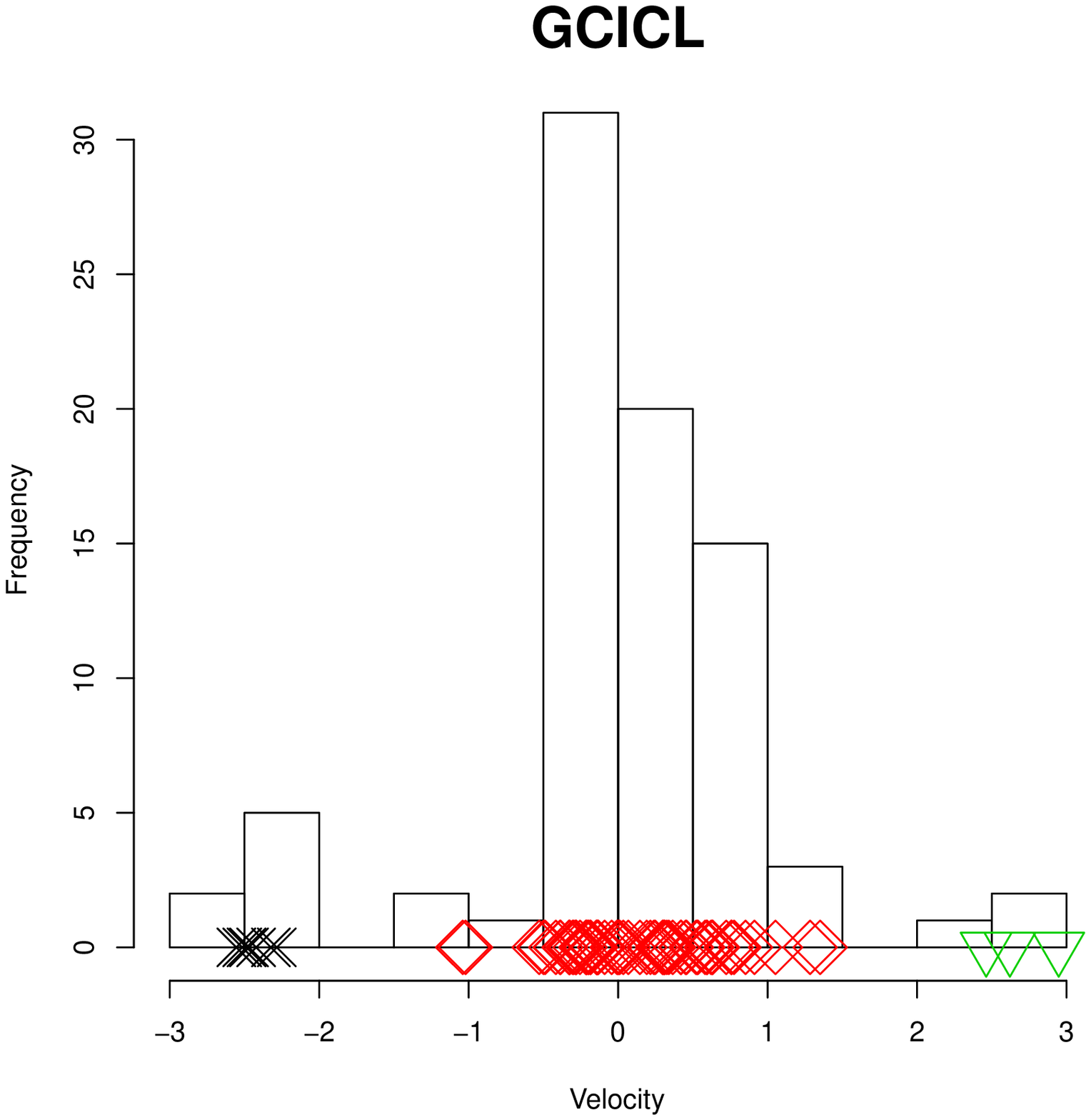}
 \caption{Clustering for the Galaxy dataset. On the left image the MBCBIC solution is represented, the MBCICL one is in the middle,
 while the GCICL is on the right. Note that the MBCICL and the GCICL configurations are equivalent.}
 \label{fig:Galaxy}
\end{figure}

The dataset considered is composed of the velocities of $82$ distant galaxies diverging from the Milky Way. The interest lies in the understanding of whether 
velocity can be used to discriminate clusters of galaxies. The data has been first analysed from a statistical point of view in \textcite{roeder1990density}.
The Greedy Combined ICL has been run for several possible choices of hyperparameters. 
The maximum number of iterations allowed and the number of reruns have been both set to $10$.
The different outcomes obtained are shown in Table \ref{table:galaxy}, while in Figure \ref{fig:Galaxy} the optimal configurations obtained using the Greedy Combined 
ICL, BIC and approximate ICL are shown.

\begin{table}[htb]
\centering
\begin{tabular}{|ccc|cc|}
\hline
\multicolumn{5}{|c|}{Galaxy dataset}\\
\hline
 $\tau$ &$\delta$& $\alpha$ & $k$ & \multicolumn{1}{c|}{$ICL_{ex}$} \\ 
\hline
0.1 &   1 & 0.5 &   3 & -109.70 \\ 
  0.1 &   1 &  10 &   2 & -120.37 \\ 
  0.1 & 0.1 & 0.5 &   3 & -105.17 \\ 
  0.1 & 0.1 &  10 &   4 & -115.76 \\ 
  0.1 & 0.01 & 0.5 &   3 & -110.11 \\ 
  0.1 & 0.01 &  10 &   3 & -117.26 \\ 
  0.01 &   1 & 0.5 &   3 & -111.51 \\ 
  0.01 &   1 &  10 &   2 & -122.53 \\ 
  0.01 & 0.1 & 0.5 &   3 & -101.85 \\ 
  0.01 & 0.1 &  10 &   3 & -114.36 \\ 
  0.01 & 0.01 & 0.5 &   3 & -103.10 \\ 
  0.01 & 0.01 &  10 &   3 & -115.61 \\ 
  0.001 &   1 & 0.5 &   3 & -114.77 \\ 
  0.001 &   1 &  10 &   2 & -124.83 \\ 
  0.001 & 0.1 & 0.5 &   3 & -104.03 \\ 
  0.001 & 0.1 &  10 &   3 & -116.54 \\ 
  0.001 & 0.01 & 0.5 &   3 & -103.48 \\ 
  0.001 & 0.01 &  10 &   4 & -118.70 \\ 
  \hline 
\end{tabular}
\caption{\textbf{Galaxy Dataset.} Best $ICL_{ex}$ values and number of groups obtained through the Greedy Combined ICL for the Galaxy dataset, for different sets of hyperparameters. 
$\gamma$ is throughout fixed to $1$ and $\mu$ to $0$ (data has been standardised).}
\label{table:galaxy}
\end{table}
The optimal number of groups is rather small with respect to other previous works (see \textcite{aitkin2001likelihood} for a summary) for all the hyperparameters' values tested.

\section{Final remarks}\label{sec:FinalRemarks}
In this paper we have shown that in a Gaussian mixture model, under the assumption of conjugate priors, an exact formula for the 
Integrated Completed Likelihood can be obtained.
A greedy algorithm, called Greedy Combined ICL, has been proposed to optimise the $ICL_{ex}$ with respect to the allocations of the observations.
In contrast to other well known methods \parencite{biernacki2000assessing}, the exact value of ICL, rather than an approximate one, is maximised. Such exact ICL differs from the approximate one
of \textcite{biernacki2000assessing} since it depends on the hyperparameters of the model, making the clustering solution sensitive to the prior distributions' choice. This fact makes perfect sense in a 
strict Bayesian context, when one only uses the observed data to update the prior information using the likelihood. However, since non-informative hyperparameters are not available for the model considered, sensitivity to priors' 
choice can become troublesome in a more objective frequentist approach. Indeed, the clustering configurations obtained through the maximization of $ICL_{ex}$ have a more subjective nature, and as such a formal comparison with other 
techniques may have very little meaning.

Two main advantages of the methodology proposed are that the greedy routine can scale 
very well with the size of the data, and that the number of groups can be inferred automatically from the final solution, since categorical allocation variables are used.
Also, since we are optimising, no issues related to the label switching problem arise, in contrast to the situation where one samples from the posterior, as in \textcite{richardson:green97}.
The optimization routine proposed stems from the Iterated Conditional Modes of \textcite{besag1986statistical} and extends similar approaches presented in 
\textcite{wyse2014inferring,come2013model}, improving efficiency and adapting to the different context.
In this paper we have extended these greedy routines introducing a combined update of groups of allocation variables, ensuring better convergence properties 
at a negligible additional computational cost. 
A possible alternative approach to the optimization problem would make use of sampling: a more informative solution could be obtained, 
essentially providing a characterization of the posterior distribution of the allocations. 

The mixture model underlying is a special case of that presented in \textcite{nobile2007bayesian}. The main difference lies in that only the symmetric case is considered here,
avoiding the dependence of the hyperparameters on the number of groups. 
However, a possible extension of our algorithm may include hyperparameters defined differently for each $K$. 

Another point to make is that many other statistical models offer the possibility to integrate out all the parameters, so that other exact formulas for the ICL are available. 
In \textcite{nobile2007bayesian} some examples are shown. 
Therefore, the same methodology introduced here can be extended to several other context and mixture frameworks.


The optimization routine introduced has been applied to a real dataset and various simulated ones.
The framework described appears to be consistent and to give meaningful clustering solutions in all the cases described. 
However, it also appears that hyperparameters can have a strong influence on the quality of the optimal solution. 
Indeed, one main drawback of the methodology exposed is that in the finite mixture framework used, pure non-informative hyperparameters are not available. 
Therefore, the user is technically forced to use informative priors, which will create unavoidable bias in the results.
Nevertheless, this limitation affects only the case where no prior information is possessed.

Other possible extensions of this work include the use of different initial configurations, which can definitely improve convergence, but also the introduction of pruning or other techniques to improve
the efficiency of the optimization, as shown in \textcite{wyse2014inferring}.
Also, the same framework described in this work can be extended to the general supervised classification case. 
Assume that we can split the data into a training set $\textbf{x}$ and a test set $\textbf{x}'$, where,
the allocations $\textbf{z}$ of the first set are known, while for the latter they are not.
So the goal is to classify the observations in the test set, i.e. find their allocations $\textbf{z}'$.
Then one can simply run the Greedy Combined ICL on the complete dataset given by the positions $y=(\textbf{x},\textbf{x}')$ 
and the allocations $\boldsymbol{\eta}=(\textbf{z},\textbf{z}')$ 
taking care of optimising only on the unknown variables, leaving the others fixed. 
Such extension could be of great interest since the method would integrate the known information on the true $K$ in a very natural way.\\


All the code used in the paper has been written in \texttt{R}, making use of the package \texttt{mclust} \parencite{fraley2006mclust} to obtain the MBCBIC and MBCICL clustering solutions. 

\section*{Acknowledgements}
The authors would like to thank Gilles Celeux for some helpful comments on an earlier draft of this paper. 
Marco Bertoletti completed this work while visiting the School of Mathematical Sciences, UCD, as part of an M.Sc. project. 
The Insight Centre for Data Analytics is supported by Science Foundation Ireland under Grant Number SFI/12/RC/2289.
Nial Friel and Riccardo Rastelli's research was also supported by a Science Foundation Ireland grant: 12/IP/1424.


\printbibliography
\end{document}